\documentclass[aps,physrev,reprint,onecolumn,groupedaddress]{revtex4-2}

\usepackage{amsmath}
\usepackage{graphicx}
\usepackage{xcolor}

\begin{document}


\title{Model of deep zonal flows in giant planets}


\author{Laura K. Currie}
\email[]{laura.currie@durham.ac.uk}
\affiliation{Department of Mathematical Sciences, Durham University, Durham DH1 3LE, UK}

\author{Chris A. Jones}
\affiliation{School of Mathematics, Statistics and Physics, Newcastle University, Newcastle upon Tyne NE1 7RU, UK}


\date{\today}

\begin{abstract}
    A mechanism by which the surface zonal flows of giant planets can be gradually attenuated with depth is explored. The zonal flow is driven by an imposed forcing in a thin layer near the surface. A meridional circulation is set up, analogous to the Ferrel-like cells observed in Jupiter's atmosphere. Acting on a stably stratified thin surface layer, the meridional flow induces a horizontal temperature anomaly which leads to a gradual reduction of the zonal winds with depth, governed by the thermal wind equation. Our model is a Boussinesq plane layer, with gravity acting parallel to the rotation axis.
    A suite of fully three-dimensional time-dependent numerical simulations has been performed to investigate the model behaviour. Below the forced stable layer, convection is occurring, typically in the form of tall thin cells. The fluctuating components of the three-dimensional flow can be driven by either the convection or the Reynolds stresses associated with the jet shear flow. When fluctuations are mainly driven by convection in the form of tall thin columns and the forcing is relatively weak, the horizontal temperature anomaly persists much deeper into the interior than when it is driven by shear flow. The zonal jets can therefore extend deep into the interior, consistent with the Juno gravity data.    
\end{abstract}


\maketitle



\section{Introduction}

\begin{figure*}
\includegraphics[scale=1]{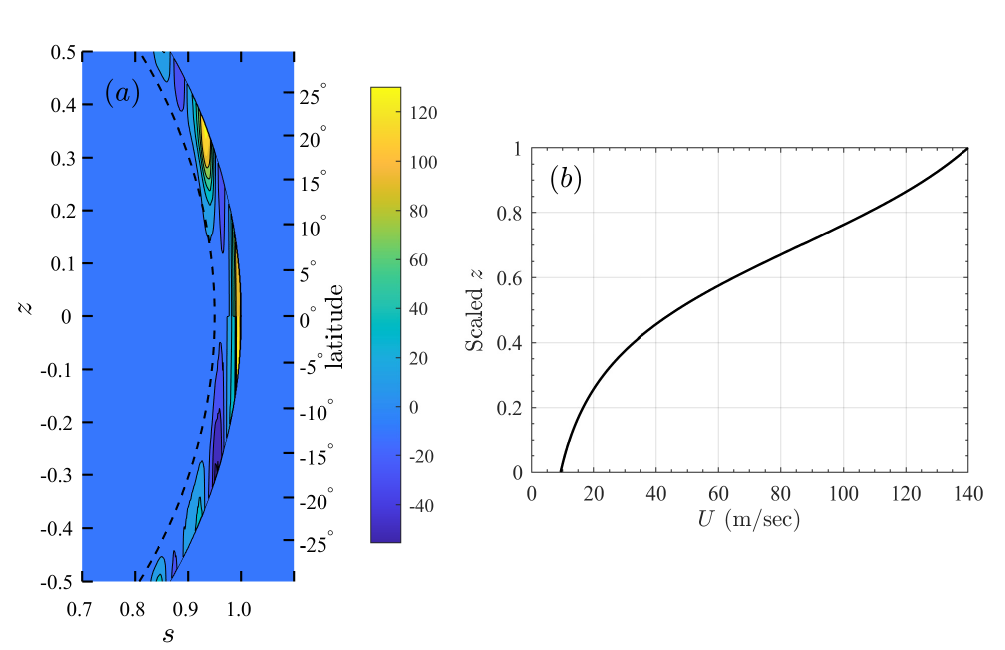} 
\caption{ (a) The deep zonal flow model of Kaspi et al. \cite{Kaspi23}. The $z$-coordinate is parallel
to the rotation axis, and the equivalent
latitude at the surface is shown (for details see Appendix A). Positive wind speeds correspond to eastward
flow in metres/sec and negative wind speeds correspond to westward flow. Only latitudes
between $\pm 30^\circ$ are shown. The dashed line is at a depth of 3500\,km below
the surface. (b) A zonal flow profile consistent with the gravity data, for the jet at latitude $21^\circ$\,N. The $z$-coordinate here is scaled so that
$z=0$ is the bottom of the layer, $z=1$ is the top of the layer, see Appendix A for details. The unscaled layer depth is 12780\,km. }
\label{figs:fig1}
\end{figure*}

It is now known that Jupiter's zonal winds extend down about 3000 km into its interior \citep{Kaspi18}. The Juno mission has given us the values of the $J_n$ coefficients of the gravity field up to degree 30 \cite{Kaspi23}. These indicate that the zonal flows are cylindrically oriented (i.e. only weakly dependent on the coordinate $z$ parallel to the rotation axis), see Fig.\,\ref{figs:fig1}(a), which is constructed from data in \cite{Kaspi23} and \citep{Jones23}. The gravity data indicates that the amplitude of the zonal flow is gradually quenched with depth, as shown in Fig.\,\ref{figs:fig1}(b), the deepest jets being those at planetocentric latitudes $21^\circ$\,N and $18^\circ$\,S which make the largest contributions to the gravity signal. While the deep zonal flow in the lower latitude region $\pm 25^\circ$ is well-constrained, uncertainty remains about the zonal flow at shallower depths, where the density is too low to affect the gravity significantly, as well as about the deep flow at higher latitudes \citep{Galanti21}. The electrical conductivity falls off rapidly above the metallic hydrogen region \citep{French2012}, which reaches to the top of the dynamo region  around 7000\,km depth \citep{TJ20}. The slow variation of the wind speed with $z$ is primarily due to temperature variations, the thermal wind effect, rather than Lorentz forces as these are very small at a depth of only $3500$\,km. 

Although the decrease of the jet speed with $z$ is gradual, it does require a variation of temperature of order $1^\circ\,K$ across the jet. This is orders of magnitude larger than the temperature variations which drive convection, and so models of rotating convection-driven jet flows in spherical shells show almost no variation of the zonal flow with $z$  \citep{Heimpel05, JK09}.
The most likely source for the temperature gradient perpendicular to $z$ is a meridional circulation driven in a stably stratified region of Jupiter. It was suggested in {\citep{Christensen20,CW24}} that there is a stable layer at a depth of approximately 2000\,km, and this can explain the observed quenching of the zonal flow in a way compatible with the Juno gravity data, see also \cite{Gastine21, Moore22}. While this is an attractive explanation of how the jets are quenched, there is as yet no explanation of why there is a stable layer at the required depth. Helium rain might lead to stable layer formation, but this occurs much deeper, near the metallic hydrogen layer \citep{Brygoo21}. The chemistry of Jupiter's interior is complex, and so the opacity is not well known, but current Jupiter models do not suggest that a stable layer could be produced at the required depth. Here we study the effect of a stable layer near the surface to see whether such a layer can also explain the rate of quenching of Jupiter's zonal jets. There is no doubt that there is a stable layer in Jupiter's atmosphere as there is in the Earth's atmosphere. The problem is that the layer is very thin, which makes it difficult numerically to incorporate into a model thousands of kilometers deep. Here we compromise by studying surface stable layers deeper than Jupiter's in order to see whether a stable layer near the surface can lead to jet quenching at depth.

There is general agreement that zonal jets can be driven by turbulence induced by convection in a rotating system \cite{Read24}. There
is less agreement about the depth at which the jets are forced. 
Showman et al. \cite{Showman06} showed that forcing in a shallow layer in the atmosphere can produce deep zonal jets.  Deep convection models e.g. \citep{Heimpel05, JK09, Heimpel22}
can produce Jupiter-like jet flows, but so can models where the turbulence is confined to the weather layer above the 3 bar level, approximately 35 km below the 1 bar surface, \citep{Schneider09}. Ingersoll et al. \cite{Ingersoll81} and Salyk {et al.} \cite{Salyk06} estimated the Reynolds stresses arising from Jupiter's eddies using Voyager data, and found them to be well-correlated with the zonal jets. They also noted that if the stresses continued downwards to the 10 bar level (100 km below the 1 bar surface) at the same level of intensity the rate of supply of energy driving the zonal flows would exceed the thermal energy flux coming out of Jupiter's interior. Since the thermal energy is the most likely ultimate source that drives the jets, this makes it likely that the Reynolds stresses fall off with depth. It is then uncertain whether the main contribution to the driving comes from the strongly driven but shallow surface layers or the much thicker but more weakly driven flows below. Possibly the driving may be a combination of deep and shallow forcing, but here we consider the end member scenario in which the primary forcing is from a shallow layer near the surface. We model this by including an imposed forcing region to represent the strong (observed) Reynolds stresses driven by small-scale unresolved eddies. Since the zonal flow alternates between being eastward and westward as the latitude varies, the imposed forcing varies in sign with latitude, consistent with the Voyager observations \cite{Ingersoll81}. 

The forcing in the zonal direction will drive an axisymmetric  meridional circulation. Evidence for these midlatitude Ferrel-like cells, gathered from Juno data, is given by Duer et al. \cite{Duer21}. Measurements of the microwave radiation strongly suggest that upwelling and downwelling
is occurring in the subcloud zone, 50-100\,km below the 1 bar surface level,  and that the Ferrel-like cells have a latitudinal extent similar to the spacing between the zonal flows. Each cell is associated with a zonal jet; in the northern hemisphere westward jets have a clockwise meridional circulation, eastward jets have an anticlockwise circulation, the opposite in the southern hemisphere. The depth of these meridional cells is not known from observations, but the cells will produce a horizontal temperature gradient as they create vertical 
motion in the stably stratified layers of the atmosphere. This horizontal temperature
gradient gives rise to a thermal wind balance, reducing the strength of the zonal flow with depth. Since in a steady state the vertical profile of the jet speed is controlled by the horizontal temperature gradient, the key to determining this profile is to understand the mechanisms by which the horizontal temperature gradient is controlled.    
Laminar thermal diffusion would reduce the horizontal temperature gradient
as the depth increases, but since thermal convection is very likely transporting the internal heat flux outwards, the diffusion will be greatly enhanced by turbulence. It is also  likely that the rotational turbulence will lead to anisotropic eddy diffusion. It is also possible shear flow instabilities coming from the jets themselves will generate turbulence at depth.

Another issue is how the azimuthal momentum balance is achieved in the jets. Whether the forcing comes from the strong turbulence in the atmosphere or the Reynolds stresses deep in the interior, in a steady state these forces must be balanced somehow. In giant planets the braking force could be magnetic Lorentz forces acting deep down. Since the currents associated with the magnetic field grow rapidly as we reach the highly conducting metallic hydrogen layer, it might be that even a small shear there could provide sufficient force to balance the driving Reynolds stresses. Another possibility is that
breaking waves at depth could provide sufficient stress to balance the driving Reynolds stresses higher up, Ingersoll et al. \cite{Ingersoll21}.
We do not address this issue here, instead invoking a no-slip boundary at the bottom to balance the forcing. This introduces unphysical shear layers at the bottom of our model, but the simplicity of this model provides a useful stepping stone before tackling the more physical MHD problem.   

Here we consider a much simplified model to explore the effect of a force applied near the top of a Boussinesq plane layer of fluid in which rotating
convection is occurring. For simplicity, we take the rotation vector parallel to gravity in the $z$ direction, and consider a force applied in the 
$x$ direction, which corresponds to the azimuthal direction in a planet, and assume the forcing varies sinusoidally in $y$. Since the giant planet jets 
occur at mid-latitudes, the rotation vector is not parallel to gravity, but here we consider the simplest geometry first. 
This problem in which a sinusoidal force is applied near the top of a plane layer of rotating convection does not seem to have been studied previously. We choose a forcing that operates at the top of the layer over a thin region, which has the advantage that a boundary layer analysis of the thin forced region is possible in the rapidly rotating limit. Our main focus here is not on how the zonal flows are driven, but instead how they vary with depth in the interior of the layer. 

As expected, we found that a stable layer near the surface is necessary to get deep zonal flows that decay gradually downwards as in Jupiter. We assume there is a thin stably stratified region lying above a convectively unstable interior, and we take the thickness of the stable layer to be comparable to the thickness of the forcing layer. Stable layers in giant planets typically have temperature gradients (entropy gradients in the compressible case)
that are much larger than the temperature (entropy) gradients in the convective regions.

In section II we describe the model we used in more detail and set up the model equations. In section III we consider the results of our simulations 
to see how different types of zonal flow can be set up. In subsections B, C and D we look at a simplified linear basic state model which is valid when the flow is only mildly forced and the convective driving is sufficiently small for the temperature to be close to its conduction state value. Since this
state can be solved using only ODEs, and in an appropriate limit an asymptotic analysis is possible, this basic linear state provides a useful check
of the numerics, and provides a framework in which the more realistic cases of strong forcing and strong convection can be understood. In subsections
III E, F and G, we consider strongly convecting models while varying the forcing and we examine when the fluctuating terms are driven mainly by the convection or mainly by the forcing. In Section IV we give a discussion and our conclusions.

\section{Model setup}

\subsection{Governing equations}

We consider a Cartesian layer (dimensions $L_x d \times L_y d \times d$) of fluid rotating about an axis aligned with gravity. The rotation vector is given by $\mathbf{\Omega} = (0, 0, \Omega)$ where $\Omega$ is the rotation rate. We take $x$ to be in the eastwards (zonal) direction, $y$ in the northwards (meridional) direction and $z$ in the upwards (radial) direction. We assume the Boussinesq approximation so that the flow is incompressible. A fixed amount of heat flux enters the domain through the bottom boundary and drives convection. In giant planets, the upper atmosphere is stably stratified because of cooling through radiative transfer. We incorporate a thin stably stratified region above the convecting region in our model via the temperature equation,
\begin{equation}
    \frac{\partial T}{\partial t} + (\mathbf{u}\cdot\nabla)T =  \nabla^2 T - \frac{Q}{H_S} \exp \left( \frac{z-1}{H_s} \right),
    \label{eq:temperature_eq}
\end{equation}
where $T$ is the temperature, $\mathbf{u}=(u,v,w)$ is the fluid velocity, $Q$ is a measure of the strength of the radiative cooling and $H_s$ is the thickness scale height of the stably stratified region. We have non-dimensionalised  this equation (and all subsequent equations) with respect to the height of the layer, $d$, and the thermal diffusion time ($d^2/\kappa$, $\kappa$ is the thermal diffusivity). Temperatures are scaled with ${\cal F}d/\kappa$, where ${\cal F}$ is the imposed heat flux entering the bottom of the layer from below.

We write the temperature as the sum of the basic state temperature profile, $T_{BS}$, and a perturbation to this state, $\theta$, so   
\begin{eqnarray}
 T &=& T_{BS}+\theta, \qquad    \frac{d^2 T_{BS}}{dz^2} = \frac{Q}{H_s} \exp  \left(\frac{z-1}{H_s} \right), \nonumber \\
T_{BS} &=& Q H_s \exp  \left(\frac{z-1}{H_s} \right) - Q z \exp \left(-\frac{1}{H_s} \right)
  -  z +T_0 - Q H_s \exp \left(-\frac{1}{H_s} \right),
  \label{eq:basicstate}
 \end{eqnarray}
where $T_0$ is the temperature at $z=0$ which we fix to be $1$ throughout. The basic state has a dimensionless temperature gradient of $-1$ at $z=0$, and since we choose $H_s \ll 1$, the second and fourth terms in the $T_{BS}$ expression are negligible, so the temperature gradient is close to $Q-1$ at the top, $z=1$. Since the buoyancy frequency of giant planet stable layers is large, $Q$ is large, so heat is conducted in at the top to allow for the sink in the heat equation. In a compressible layer it would only be the potential temperature that increases upwards, not the actual temperature, so our Boussinesq temperature should be thought of as a potential temperature in a real giant planet. The height at which the basic state temperature gradient changes sign is $z_{neu} \approx 1 + H_s \ln (1/Q)$, which indicates how thick our stable layer is. Note though that although $T_{BS}$ is fixed in time in our model, when the system convects strongly, the location where the full temperature $T$ changes sign might not be at $z_{neu}$. Illustrative profiles of $T_{BS}$ for a case without a stable layer ($Q=0$) and a case with a stable layer ($Q=200, H_s=0.02831$) are shown in Fig. \ref{figs:fig2}(a). 

The dimensionless governing equations are:
\begin{equation}
    \frac{\partial \mathbf{u}}{\partial t} + (\mathbf{u}\cdot\nabla)\mathbf{u} +\nabla p + Ta^{\frac{1}{2}}Pr \mathbf{\hat z} \times \mathbf{u}  -Ra_F Pr \theta {\bf \hat z} =  Pr\nabla^2 \mathbf{u} +\mathbf{F},
    \label{eq:momentum_eq}
\end{equation}
\begin{equation}
    \nabla\cdot\mathbf{u} = 0,
    \label{eq:continuity_eq}
\end{equation}
\begin{equation}
    \frac{\partial \theta}{\partial t} + (\mathbf{u}\cdot\nabla)\theta +w\frac{dT_{BS}}{dz}=  \nabla^2 \theta,
    \label{eq:theta_eq}
\end{equation}
where $p$ is the pressure. We have also introduced the following non-dimensional parameters:
\begin{equation}
    Pr=\frac{\nu}{\kappa}, \quad Ta=\frac{4\Omega^2d^4}{\nu^2}, \quad Ra_F = \frac{\alpha g {\cal F} d^4}{\nu\kappa^2},
    \label{eq:parameters_eq}
\end{equation}
where $\nu$ is the kinematic viscosity, $\alpha$ is the thermal expansion coefficient and $g$ is the acceleration due to gravity.

The jets in giant planets are likely driven mainly by turbulent Reynolds stresses in the thin weather layer, and as this small scale turbulence is difficult to compute numerically, we model it by an imposed forcing in the $x$-direction confined to a relatively thin layer near the top. In particular, we take the forcing $\mathbf{F}$ in equation (\ref{eq:momentum_eq}) to be 
\begin{equation}
    \mathbf{F} = F\cos(k_y y)\exp \left(\frac{z-1}{H_f} \right) \hat{\mathbf{x}},
         \label{eq:Fdef}
\end{equation}
where $F$ is the magnitude of the forcing, $k_y$ is the wavenumber of the sinusoidal variation in $y$ and $H_f$ is the length scale over which the forcing operates.

The components of (\ref{eq:momentum_eq}) for the velocity $(u,v,w)$ are
\begin{eqnarray}
 \frac{\partial u}{\partial t} + (\mathbf{u}\cdot\nabla) u +\frac{\partial p}{\partial x} - Ta^{\frac{1}{2}}Pr v   &=&  Pr\nabla^2 u 
  + F\cos(k_y y)\exp \left(\frac{z-1}{H_f} \right), \label{eq:xcomponent}\\
  \frac{\partial v}{\partial t} + (\mathbf{u}\cdot\nabla) v +\frac{\partial p}{\partial y} + Ta^{\frac{1}{2}}Pr u   &=&  Pr\nabla^2 v,  \label{eq:ycomponent} \\
  \frac{\partial w}{\partial t} + (\mathbf{u}\cdot\nabla) w +\frac{\partial p}{\partial z} - Ra_F Pr \theta   &=&  Pr\nabla^2 w \label{eq:zcomponent}.
\end{eqnarray}

At the bottom boundary, we choose a no-slip boundary condition so that the flow is brought to rest at $z=0$. In a giant planet, it is the magnetic field which will ensure that the zonal flow is brought down to the small values in the dynamo region, but hopefully modelling this as a no-slip boundary is a crude approximation to the more sophisticated MHD process. At the top boundary, we impose a stress-free boundary condition, so that the forcing can drive a sinusoidal zonal flow at the top. Both boundaries are assumed impermeable. Together with the fixed flux temperature condition this gives the boundary conditions as:
\begin{align}
    u &= v = w = \frac {\partial \theta}{\partial z} = 0 \quad \text{on } z=0, \label{bcs_bottom}\\
    \frac{\partial u}{\partial z}  &= \frac{\partial v}{\partial z}  = w = \frac {\partial \theta}{\partial z} = 0 \quad \text{on } z=1.
    \label{bcs_top}
\end{align}
 The horizontal boundaries are taken to be periodic. 
 The simulations we present all have $Pr=1$, $L_x=L_y=4$ and $k_y=\pi/2$, so one full wavelength of the zonal flow fits into the periodic box. 
In Table \ref{table:cases3} the parameters used in the various runs are specified.

\subsection{Numerical method and output quantities}
We solve equations (\ref{eq:momentum_eq})-(\ref{eq:theta_eq}) subject to boundary conditions (\ref{bcs_bottom}) and (\ref{bcs_top}) using the open source pseudo-spectral solver Dedalus \cite{Burns20}. All fields are expanded in a Fourier basis of $N_x=N_y$ modes in the $x$ and $y$ directions, and in a Chebyshev basis of $N_z$ modes in the $z$ direction. For the majority of simulations, we use a resolution of $N_x=N_y=N_z=128$ spectral modes in each spatial direction, however, a higher resolution of $[N_x\times N_y \times N_z]=[256 \times 256 \times 128]$ modes is used for runs at $Ta=10^9$. A $3/2$ dealiasing rule is applied in all directions such that nonlinearities are evaluated in physical space on a grid of size $3/2 \times [N_x \times N_y \times N_z]$.  A 2nd order Runge-Kutta implicit-explicit time stepping scheme is used; the linear terms are time stepped implicitly and the nonlinear terms explicitly. The simulations are initialised with either a small-amplitude temperature perturbation or an existing solution and are evolved until a statistically stationary regime is clearly obtained.  

If there is no forcing and no stable layer, i.e. the parameters $F$ and $Q$ are both zero, our problem reduces to classical Boussinesq rotating convection in a plane layer, which is a very well-studied problem, both with laboratory experiments and using numerical simulations, see e.g. \cite{Ecke23}. 
With $F=0$ and $Q$ non-zero there is a critical Rayleigh number, $Ra_{crit}$, and wavenumber, $k_{crit}$, for the onset of convection for the motionless purely thermal conduction state. In Table \ref{table:cases3} $Ra_{crit}$ and $k_{crit}$ are listed, computed by solving the linear eigenvalue problem numerically (details are given in section \ref{sec:linear} below). Note, we take $k_y=\pi/2$, which is much less than the values of $k_{crit}$ associated with the $F=0$ linear stability problem.  If $F \ne 0$ there is a steady two-dimensional 
flow even if $Ra_{F} = 0$, so determining the critical $Ra_{F}$ for the onset of convection is computationally demanding, though if $F$ is small it will be close to $Ra_{crit}$.

While snapshots of the simulation solutions give useful insight into the dynamics, our main focus is on the time-average of the zonal flow, which in our geometry is the $x$-averaged flow in the $x$-direction. We therefore define the mean and fluctuating parts of the variables by taking the average over $x$ and $t$, so for example
\begin{equation} \label{meandefn}
{\bar \theta} = \langle\theta\rangle  = \frac{1}{\tau L_x} \int_0^\tau \int_0^{L_x} \theta \, dx dt, \qquad \theta' = \theta - {\bar \theta},
\end{equation}
where $\tau$ is the total integration time and similarly for $(\bar u,\bar v,\bar w)$  for the velocity components $(u,v,w)$. 
In the simulations, the mean (barred) variables are calculated by averaging from a time $\tau_1$ to the time at the end of the simulation $\tau_2=\tau$, e.g., 
\begin{equation}
    \bar u = \frac{1}{\tau_2 - \tau_1} \int_{\tau_1}^{\tau_2} \langle u \rangle_x \, dt.
\end{equation}
A suitable value of $\tau_1$ was determined by examining running averages, and choosing a $\tau_1$ such that initial transient behaviour was not strongly influencing the mean quantities.  $\tau_1=0.5$ was used for most cases, except for those with $Ra_F=0$ where the initial transient lasts longer.  The values of $\tau_2$ used for each run are given in Table \ref{table:cases3}.

Occasionally, we examine snapshots of 
an $x$-averaged quantity; for example  $\langle w\rangle _x$  denotes the vertical velocity averaged over $x$ but at a particular $y$, $z$ and $t$.  

We also give some output Rossby numbers in Table \ref{table:cases3}. We define
\begin{equation}
    Ro_{\bar u} = \frac{1}{Ta^{\frac{1}{2}}Pr} \left[  \frac{1}{L_y} \int_{0}^{1} \int_{0}^{L_y} {\bar u}^2 \,dy\,dz  \right]^{1/2}, \ 
    Ro_{{\bar v} {\bar w}} = \frac{1}{Ta^{\frac{1}{2}}Pr} \left[  \frac{1}{L_y} \int_{0}^{1} \int_{0}^{L_y} ({\bar v}^2 + {\bar w}^2) \,dy\,dz  \right]^{1/2},
    \label{Ro_ubar_def}
\end{equation}
so $Ro_{\bar u}$ is the Rossby number based on the mean zonal flow velocity, and  $Ro_{\bar {v} \bar{w}}$ is the Rossby number for the mean meridional circulation, which is normally weaker than the zonal flow. $Ro_{vw}$ evaluates the fluctuation Rossby number and is defined as
\begin{equation}
    Ro_{vw} = \frac{1}{Ta^{\frac{1}{2}}Pr} \left[  \frac{1}{L_xL_y} \int_{0}^{1} \int_{0}^{L_y} \int_{0}^{L_x} ({v}^2 + {w}^2) \,dx\,dy\,dz  \right]^{1/2}.
    \label{Ro_vw_def}
\end{equation}
 The value we give is calculated from one snapshot near the end of the run, rather than a time average, though several values from different snapshots were computed to check that these values are typical.  This captures the magnitude of the fluctuating convective parts of the flow. Because these fluctuating parts average out over time during the run, $\bar v$ and $\bar w$ are much smaller than the instantaneous $v$ and $w$,
so $Ro_{\bar {v} \bar{w}} \ll Ro_{vw}$ if convection is active.

\subsection{Mean equations}
The mean quantities defined in (\ref{meandefn}) must satisfy equations formed by taking the means $\langle\rangle $ of equations (\ref{eq:xcomponent})-(\ref{eq:zcomponent}), (\ref{eq:continuity_eq}) and (\ref{eq:theta_eq}):
\begin{equation}
     (\mathbf{{\bar u}}\cdot\nabla){\bar u}  + \left\langle \frac{\partial}{\partial y} (v'u') + \frac{\partial}{\partial z} (w'u') \right\rangle  -
     Ta^{\frac{1}{2}}Pr {\bar v} =  Pr\nabla^2 {\bar u}  + F\cos(k_y y)\exp \left(\frac{z-1}{H_f} \right),
    \label{mean_xmomentum_eq}
\end{equation}

\begin{equation}
    (\mathbf{{\bar u}}\cdot\nabla){\bar v} +  \left\langle \frac{\partial}{\partial y} (v'^2) + \frac{\partial}{\partial z} (w'v') \right\rangle   + 
    \frac{\partial {\bar p}}{\partial y} +Ta^{\frac{1}{2}}Pr {\bar u} =  Pr\nabla^2 {\bar v},
    \label{mean_ymomentum_eq}
\end{equation}

\begin{equation}
    (\mathbf{{\bar u}}\cdot\nabla){\bar w}  +  \left\langle \frac{\partial}{\partial y} (v'w') + \frac{\partial}{\partial z} (w'^2) \right\rangle  +
   \frac{\partial {\bar p}}{\partial z}  - Ra_FPr {\bar \theta} =  Pr\nabla^2 {\bar w},
    \label{mean_zmomentum_eq}
\end{equation}
\begin{equation}
\frac{\partial {\bar v}}{\partial y} + \frac{\partial {\bar w}}{\partial z} =0,
    \label{mean_continuity_eq}
\end{equation}

\begin{equation}
    (\mathbf{{\bar u}}\cdot\nabla){\bar \theta} + \left\langle \frac{\partial}{\partial y} (v'\theta') + \frac{\partial}{\partial z} (w' \theta') \right\rangle  +
    {\bar w} \frac{dT_{BS}}{dz} =  \nabla^2 {\bar \theta}.
    \label{mean_temperature_eq}
\end{equation}
In these equations terms involving $\partial / \partial x$ or $\partial / \partial t$ have vanished because of the averaging. 
The primed terms in the equations of motion are the Reynolds stresses, which are primarily caused by the turbulent flow set up by the
convection. The primed terms in the heat equation are turbulent advection terms, which can lead to eddy diffusion of the temperature field.


\section{Simulation results}

\subsection{Zonal flow}
\begin{figure*}
\includegraphics[scale=1]{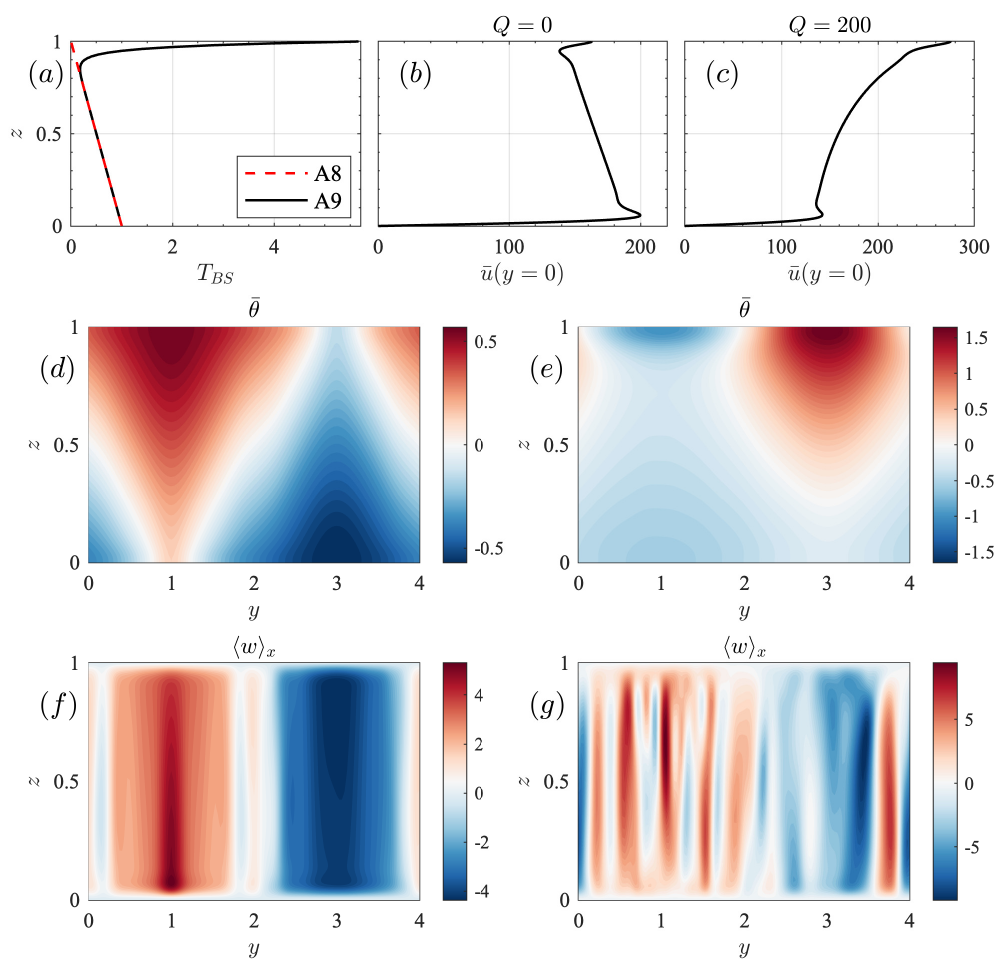} 
\caption{Simulation results for run A8 with no stable layer (Q=0) and run A9 with a strong stable layer (Q=200). (a) shows the basic state temperature as given by equation (\ref{eq:basicstate}) for the two runs. (b) and (c) show the mean zonal flow, $\bar u$, as a function of $z$ at $y=0$. (d) and (e) show the mean temperature perturbation $\bar \theta$. Note that in the two runs the sign of $\bar \theta$ is reversed in the $y$-direction. Panels (f) and (g) show the $x$-averaged vertical velocity at a snapshot in time, $\langle w \rangle_x$. These are broadly similar, but the columnar structure of the convection is particularly visible in run A9 (g).
}
\label{figs:fig2}
\end{figure*}

We first consider the effect of the thin stable region at the top of the layer. Figure \ref{figs:fig2} shows outputs from two simulations, one with no stable layer (run A8) and one with a stable layer (run A9). Fig.\,\ref{figs:fig2}(a) shows the basic state temperature profile, $T_{BS}$, in the two simulations. Fig.\,\ref{figs:fig2}(b) shows the zonal flow ${\bar u}$, which is averaged over $x$ and $t$,
evaluated at $y=0$ where the forcing has its maximum positive (eastward) value for run A8 which has no stable layer. Fig.\,\ref{figs:fig2}(c) shows the same quantity for a case with a stable layer 
at the top (run A9). These can be compared with Fig.\,\ref{figs:fig1}(b), which shows the decline of the jet speed with depth derived from Juno data. Both the zonal flows derived from the simulations have boundary layers near $z=0$ and $z=1$ which are not present in Fig.\,\ref{figs:fig1}(b). The bottom boundary layer is an artefact of the no-slip
boundary, and will not be present if the flow is brought to rest by a magnetic field rather than a viscous layer. The Jupiter zonal flow data is taken from a specific level near the 1 bar depth and it is not known whether there is a boundary layer at the top of Jupiter's interior: if there is, it will be too close to the surface to be detectable
by gravity measurements. Even if there is, our model of the forcing is quite crude, and so the main focus here is on how well the drop-off of the zonal flow in the interior corresponds with the Juno data results. 

\begin{figure*}
\includegraphics[scale=1]{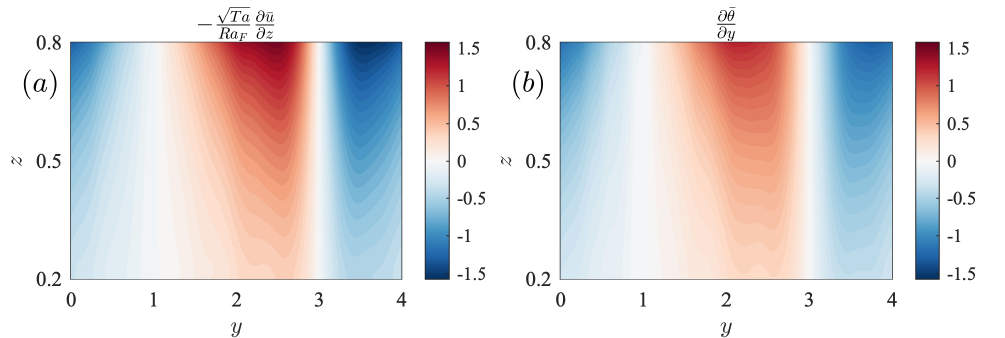} 
\caption{ Approximate thermal wind balance for run A9. The two terms
 in equation (\ref{thermal_wind_eq}) are shown to be in fairly good agreement, so that the inertial and viscous terms in equation (\ref{full_thermal_wind_eq}) are negligible in this case, and $\epsilon_{TW} = 0.098$ only.}
 \label{figs:fig3}
\end{figure*}

In the boundary layers near $z=0$ and $z=1$, the equations simplify when $Ta$ is large. In Appendix B we give the details of the solution of the boundary layer equations in both the top and bottom boundary layers. We take the forcing to have magnitude not more than $F \sim O(Ta^{3/4})$ and assume the Rayleigh number $Ra_F$ satisfies $Ra_F \ll Ta$. Since the onset of convection occurs when $Ra_F \sim Ta^{2/3}$, $Ra_F \ll Ta$ still allows vigorous convection 
to occur in the interior, but buoyancy is small in the boundary layers compared with the other terms there. A forcing $F \sim O(Ta^{3/4})$ is not so large that the inertial terms are significant in the boundary layers, though they are generally important in the interior. In Appendix B we assume the inertial terms and the buoyancy term are negligible in the boundary layers, which allows the boundary layer equations to be
solved analytically. 

It is clear from Fig.\,\ref{figs:fig2}(b) and (c) that without the stable layer, the slope of the zonal flow with $z$ has the wrong sign, but with the stable layer the slope has the correct sign. This can be understood in terms of a forced Ekman boundary layer theory (see Appendix B). The $z$-component of the vorticity in the interior is determined by
the $y$ gradient of the zonal flow $\omega_z= - \partial u / \partial y$, and at $y=3$ $\omega_z<0$ near the top boundary,
see equation (\ref{AppB_tildedef_eq}). From Appendix B we note
that the magnitude of the $z$-vorticity in the forced Ekman layer decreases as we go down from $z=1$ into the interior. In rotating systems, negative vorticity 
corresponds to increased pressure, so the pressure at $y=3$ decreases as we go into the interior, so fluid is pumped downwards at $y=3$ 
(see Figs.\,\ref{figs:fig2}(f) and (g)) and similarly sucked upwards at $y=1$ where the positive $z$ vorticity is a maximum at $z=1$.
 This suction and pumping applies whether there is a stable layer or not. With no stable layer, the suction drags warm fluid up into
 the boundary layer at $y=1$, but pumps cold fluid into the boundary layer at $y=3$, see Fig.\,\ref{figs:fig2}(d). With a stable layer,
 the temperature gradient in the $z$-direction at the top is reversed,  and so now warm fluid is pumped down at $y=3$ while cooler fluid is sucked up at $y=1$,
 see Fig.\,\ref{figs:fig2}(e). This process sets up a horizontal temperature variation, which is approximately sinusoidal as the zonal flow forcing is sinusoidal. We refer to this temperature variation set up in the surface layers as the horizontal temperature anomaly as it occurs in almost all our simulations, and it plays an important role in determining how deep the zonal flows penetrate.

In many of the simulations, the $x$-component of the curl of the mean momentum equation:
 \begin{eqnarray}
    \underbrace{ ({\bf \bar u}\cdot \nabla){\bar \omega_x} +
    \left\langle ({\bf u'} \cdot \nabla ) \omega'_x - ({\boldsymbol \omega'} 
    \cdot \nabla) u' \right\rangle }_{\text{inertial}}
    -\underbrace{Pr \nabla^2 {\bar \omega_x} }_{\text{viscous}} =  Ta^{1/2} Pr \frac{\partial {\bar u}}{\partial  z} +Ra_F Pr \frac{\partial {\bar \theta}}{\partial y},
    \label{full_thermal_wind_eq}
 \end{eqnarray}
 is dominated by thermal wind balance in the interior, so at large $Ta$ and $Ra_F$ the terms on the left-hand-side of (\ref{full_thermal_wind_eq}) are negligible compared to the terms on the right-hand-side and we have approximate thermal wind balance 
\begin{equation}
    \frac{\partial {\bar\theta}}{\partial y} \approx -\frac{Ta^{\frac{1}{2}}}{Ra_F}\frac{\partial {\bar u}}{\partial z},
    \label{thermal_wind_eq}
\end{equation}
see e.g. Fig.\,\ref{figs:fig3}. From Fig.\,\ref{figs:fig2}(e) we see that in the stable layer case $\partial {\bar \theta} / \partial y <0, $
at $y=0$, so $\partial {\bar u} / \partial z >0 $, i.e. the zonal flow strength decays with depth as in Fig.\,\ref{figs:fig1}(b), the Jupiter case,
but from Fig.\,\ref{figs:fig2}(d)  with no stable layer $\partial {\bar \theta} / \partial y >0, $
at $y=0$, so the zonal flow increases with depth as in Fig.\,\ref{figs:fig2}(b). Note however, if the forcing $F$ is strong, or $Ra_F$ is too large, the inertial terms in equation (\ref{full_thermal_wind_eq}) may become significant, so not all our simulations had perfect thermal wind balance. 

A measure of how well the thermal wind balance in the interior between $z=0.2$ and $z=0.8$ is satisfied is given by

\begin{eqnarray}
\label{eps_TW_def}
    \epsilon_{TW} &=& \frac{I_1}{I_2}, \qquad  I_1 = \left[ \frac{1}{0.6 L_y}  \int_{0.2}^{0.8} \int_0^{L_y} \left(  \frac{Ta^{1/2}}{Ra_F}  \frac{\partial {\bar u}}{\partial z} +
    \frac{\partial {\bar \theta}}{\partial y} \right)^2 \,dy\,dz, \right]^{1/2}\\ \nonumber
     I_2 &=& \left[ \frac{1}{0.6 L_y} \int_{0.2}^{0.8} \int_0^{L_y}\left(  \frac{Ta^{1/2}}{Ra_F} \frac{\partial {\bar u}}{\partial z}  \right)^2 \,dy\,dz \right]^{1/2}
       + \left[ \frac{1}{0.6 L_y} \int_{0.2}^{0.8} \int_0^{L_y} \left(   \frac{\partial {\bar \theta}}{\partial y}  \right)^2 \,dy\,dz \right]^{1/2},
\end{eqnarray}
so $I_1$ is small if the thermal wind equation is nearly satisfied, hence small $\epsilon_{TW}$ means that the thermal wind balance is accurately satisfied. Thermal wind balance is not expected to hold near the boundaries where the viscous term are significant. In Fig.\,\ref{figs:fig3} we compare the terms in the thermal wind equation (\ref{thermal_wind_eq}) for run A9. We can see
that the two terms are approximately equal in the interior, and $\epsilon_{TW}=0.098$ only, confirming the approximate balance. Values of 
$\epsilon_{TW}$ for each run are listed in Table \ref{table:cases3}. Long time averages were usually required before $\epsilon_{TW}$ settled down to its final value, and the larger $Ta$, the longer the time average needed to be.

\begin{figure*}
\includegraphics[scale=0.9]{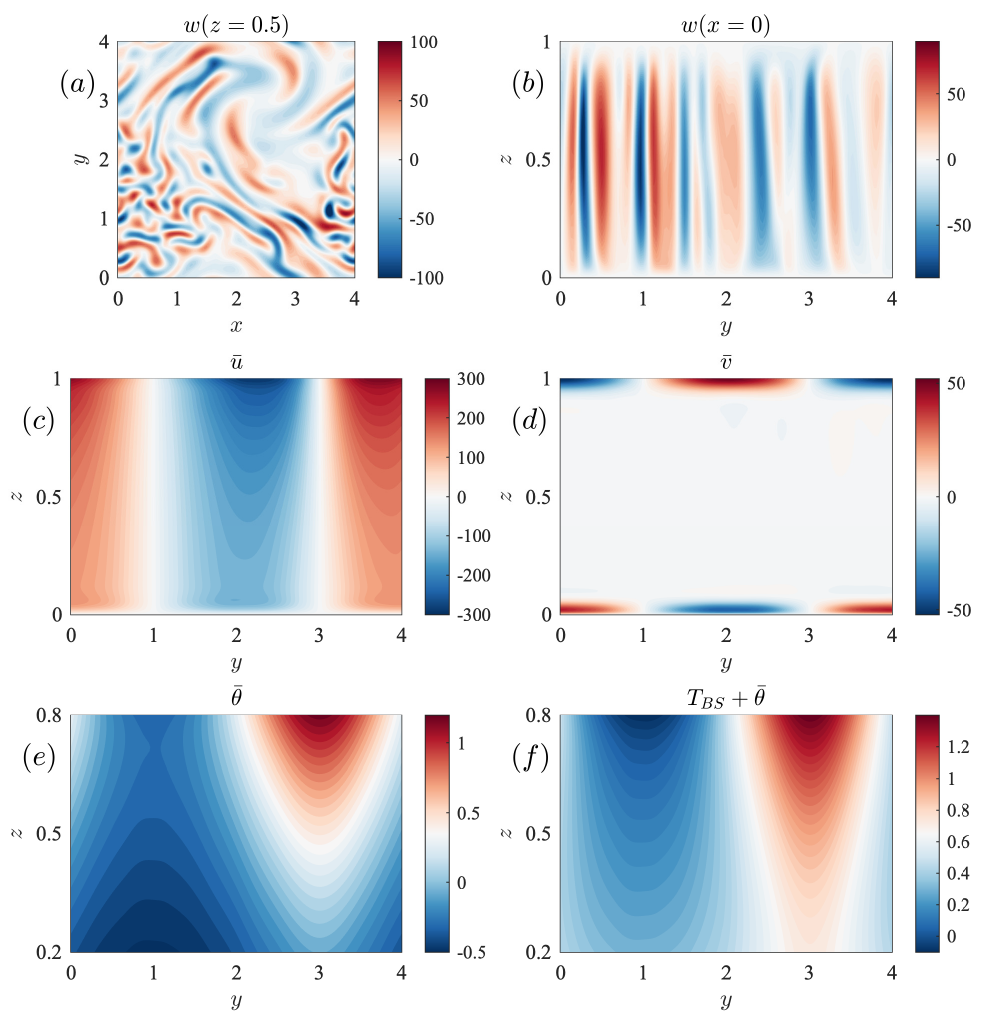} 
\caption{Further results from run A9. (a) shows a  snapshot of the vertical velocity at the  horizontal mid-plane, $z=0.5$. (b) shows a snapshot of the vertical velocity in the $y-z$ plane at $x=0$. The columnar
nature of the flow is evident here. (c) the mean zonal flow, $\bar u$, consistent with Fig.\,\ref{figs:fig3}(a). (d) the $y$-component of the mean flow, $\bar v$ which
is much stronger in the boundary layers than in the interior. (e) The mean temperature 
perturbation, $\bar \theta$ in the region $0.2 < z < 0.8$ (see Fig.\,\ref{figs:fig2}{e} for the full version) showing how the temperature perturbation decays with depth. ${\bar \theta}_{vr}=0.27$ for this run.
(f) The full mean temperature, $T_{BS}+\bar \theta$, in the region $0.2 < z < 0.8$.}
\label{figs:fig4}
\end{figure*}

In Figs.\,\ref{figs:fig2}(b) and (c) we can see that $\bar u$ varies smoothly in the interior, but jumps in the top and bottom boundary layers. In Table \ref{table:cases3} we
give for each run the jump in $\bar u$ across the top boundary layer, $\Delta {\bar u}_{top}$, across the bottom boundary layer, $\Delta {\bar u}_{bot}$, and
across the interior, $\Delta {\bar u}_{int}$. The jump $\Delta { \bar u}_{top}$ is calculated by fitting a parabola to the computed $\bar u$ in the interior just
below the boundary layer and extrapolating the parabolic fit to $z=1$. The points used for the parabolic fit were $z=[0.775,0.825,0.875]$ for $Ta=10^7$, $z=[0.825,0.875,0.925]$ for $Ta=10^8$ and $z=[0.875,0.925,0.975]$ for $Ta=10^9$. This gives an estimated ${\bar u}_{top}^{est}$ which is the value ${\bar u}$ would have at
$z=1$ if there were no boundary layer. Then $\Delta {\bar u}_{top} = {\bar u}(z=1) - {\bar u}_{top}^{est} $ where ${\bar u}(z=1)$ is the computed $\bar u$ with the boundary layer present. This procedure gave values of $\Delta {\bar u}_{top}$
in good agreement with the jump in $\bar u$ evaluated by using the forced Ekman layer theory results in Appendix B, equation (\ref{Deltauestimate_eq}), indicating that the terms omitted in that analysis do not play a crucial role in
the top boundary layer. As an example, run A9 has $H_f= (4/Ta)^{1/4}$ and $F=100 Ta^{1/2}$, which makes $\Delta u = 20 \cos (k_y y)$, so at $y=0$, $\Delta u = 20$ according to the boundary layer theory of Appendix B, which can be compared with 
$\Delta {\bar u}_{top} = 20.8$ in Table \ref{table:cases3}. At $z=1$, the boundary layer theory gives ${\bar v} = - 60 \cos (k_y y)$, see equation (\ref{AppB_vsol_eq}), which can be compared with Fig.\,\ref{figs:fig4}(d). $\Delta {\bar u}_{bot}$ was calculated similarly, extrapolating a parabolic fit to $\bar u$ just above the bottom boundary layer (using $z=[0.125,0.175,0.225]$ for $Ta=10^7$, $z=[0.075,0.125,0.175]$ for $Ta=10^8$ and $z=[0.025,0.075,0.125]$ for $Ta=10^9$) and extrapolating to get ${\bar u}_{bot}^{est}$ at $z=0$. Then $\Delta {\bar u}_{bot} =  {\bar u}_{bot}^{est} - {\bar u}(z=0)$, and $\Delta {\bar u}_{int} =  {\bar u}_{top}^{est} - {\bar u}_{bot}^{est}$ gives the jump in $\bar u$ across the interior. From Table \ref{table:cases3} we see that increasing the forcing $F$ increases the jump across the top boundary  layer
$\Delta {\bar u}_{top}$ while increasing $Ra_F$ increases the thermal wind, and hence the jump across the interior $\Delta { \bar u}_{int}$.

Another quantity useful for interpreting the results is a measure of how deeply the horizontal temperature anomaly at the top boundary penetrates into the interior. We define the temperature variation ratio, $\bar \theta_{vr}$, as
\begin{eqnarray}
\label{theta_vr_def}
   \bar \theta{_{vr}} = \frac{T_1}{T_2}, \quad T_1= \frac{1}{L_y} \int_0^{L_y} {\bar \theta}(z=0.2) \sin (k_y y) \,dy,
   \quad  T_2= \frac{1}{L_y} \int_0^{L_y} {\bar \theta}(z=0.8) \sin (k_y y) \,dy.
\end{eqnarray}
This measures how strong the $\sin (k_y y)$ component of the temperature fluctuation is at the bottom of the interior compared to the value near the top of the interior. A value of $\bar\theta_{vr}$ close to one indicates the sinusoidal variation of $\bar \theta$ continues deep into the interior, while a smaller value indicates the horizontal temperature variation has been evened out to some degree towards the bottom of the layer. The values $z=0.2$ and $z=0.8$ were chosen because in all runs they lie well clear of the boundary layers. The zonal flow induces a vertical velocity with this sinusoidal variation, and the stable stratification converts this into a sinusoidal temperature anomaly. In Figs.\,\ref{figs:fig4}(a) and (b), which are for run A9, we see that the flow is fully three-dimensional, with a tall thin columnar structure. The flow is strongly time-dependent, with columns moving around and disappearing, while new columns continually emerge. It is only when we average over $x$ and $t$ that the persistent mean quantities emerge. In Fig.\,\ref{figs:fig4}(c) we see the gradual decay of the zonal flow with depth, consistent with the thermal wind term shown in Fig.\,\ref{figs:fig3}(a). Fig.\,\ref{figs:fig4}(d) shows that the meridional flow $\bar v$ is much smaller in the interior than in the boundary layers. Figs.\,\ref{figs:fig4}(e) and (f) show the temperature perturbation $\bar \theta$ and the full temperature $T_{BS} +\bar \theta$
in the interior, showing the decay of the horizontal temperature fluctuation with depth.
For run A9, $\bar \theta_{vr}= 0.27$, so the turbulence and conduction in the interior have mostly evened out the horizontal temperature variation at the bottom of the layer. If thermal wind balance holds to a reasonable approximation, this determines how deep the zonal flows penetrate. We will see below that deep penetration occurs
in some runs ($\bar \theta_{vr}$ close to 1) while it is quite small in other runs.

%
\subsection{The basic linear state model}\label{sec:linear}

To get some insight into the behaviour of our problem, we first look here at a simplified model, dropping the nonlinear inertia terms in the equation of motion, and the nonlinear
advection term in the temperature equation. This leads to a one-dimensional system of forced linear equations which can be solved easily using MATLAB. Note that when $F \ne 0$ these linear solutions are not an exact basic state in the way that the motionless state before the onset of convection is referred to as a basic state. A similar linearised approach was used in  spherical geometry by \cite{Christensen20}. When $ F \ne 0$ the fluid always moves, whatever the value of $Ra_F$, so the nonlinear terms we are omitting will always be non-zero. However, when the forcing is small (in a sense to be specified below) and the convection is not strongly supercritical, we expect these nonlinear terms to be small, so our basic linear state model should be a reasonable approximation of the true state.  It is also possible to do some asymptotic analysis on this system. We then consider the circumstances under which dropping the inertial terms and the advection terms might be valid.

We start with the full equations (3) to (5), and drop the time dependent term and the inertial term in (3) and the time dependent term and the advection term in (5). We look for a steady solution independent of $x$ to get
\begin{equation}
    \nabla p + Ta^{\frac{1}{2}}Pr \hat {\bf z} \times \mathbf{u} -Ra_F Pr \theta \hat {\bf z} =  Pr\nabla^2 \mathbf{u} +\mathbf{F},
    \label{momentum_eq_lin}
\end{equation}
\begin{equation}
    w\left\{ Q \exp \left(\frac{z-1}{H_s}\right)  - Q \exp \left(-\frac{1}{H_s}\right) -1 \right\} =  \nabla^2 \theta.
    \label{theta_eq_lin}
\end{equation}
Taking $\hat {\bf z} \cdot \nabla \times$ of (\ref{momentum_eq_lin}) gives
\begin{equation}
   -Ta^{1/2} Pr \frac{\partial w}{\partial z} = Pr \left(  \frac{\partial^2 }{\partial y^2} + \frac{\partial^2}
   {\partial z^2} \right) \omega_z + F k_y \sin(k_y y) \exp \left( \frac{z-1}{H_f} \right) ,
   \label{zcurl_eq_lin}
\end{equation}
where $\omega_z=- \partial u / \partial y$
and taking $\hat {\bf z} \cdot \nabla \times \nabla \times $ of (\ref{momentum_eq_lin}) gives
\begin{equation}
    Ta^{1/2}  \frac{\partial \omega_z}{\partial z} =  \left( \frac{\partial^2 }{\partial y^2} + \frac{\partial^2 }{\partial z^2} \right)^2 w + Ra_F \frac{\partial^2  \theta}{\partial y^2} .
    \label{zdoublecurl_eq_lin}
\end{equation}
Since this steady system is linear, we can set
\begin{equation}
  u= {\tilde u(z)} \cos(k_y y), \quad v= {\tilde v(z)} \cos(k_y y), \quad w= {\tilde w(z)} \sin(k_y y), \quad 
 \theta= {\tilde \theta(z)} \sin{}(k_y y), \quad \omega_z= {\tilde \omega_z(z)} \sin(k_y y),  
     \label{tildedef_eq_lin}
\end{equation}
where the sinusoidal behaviour in $y$ is determined by our choice of zonal flow forcing.
Then we have
\begin{equation}
   -Ta^{1/2} Pr \frac{d \tilde w}{d z} = Pr \left(  \frac{d^2 }{d z^2} - k_y^2 \right) {\tilde \omega}_z + F k_y \exp \left( \frac{z-1}{H_f} \right) ,
    \label{lin_ord_vort}
\end{equation}
\begin{equation}
    Ta^{1/2}  \frac{d {\tilde \omega}_z}{d z} =  \left( \frac{d^2 }{d z^2} - k_y^2 \right)^2 \tilde w - k_y^2 Ra_F \tilde \theta ,
    \label{lin_ord_omz}
\end{equation}
\begin{equation} 
\tilde w \left\{ Q \exp \left(\frac{z-1}{H_s}\right) - Q \exp \left( -\frac{1}{H_s} \right)  -1 \right\} = \left( \frac{d^2 }{d z^2} - k_y^2 \right) \tilde \theta .
 \label{lin_ord_temp}
\end{equation}
These are a set of ordinary differential equations, which we solve using MATLAB. Letting
\begin{eqnarray}
  &\tilde w&= y_1, \quad \frac{d \tilde w}{dz} =y_2, \quad \frac{d^2 \tilde w}{dz^2} =y_3, \quad \frac{d^3 \tilde w}{dz^3} =y_4, \nonumber \\
 &{\tilde  \omega}_z&= y_5,  \quad \frac{{d \tilde \omega}_z}{dz} = y_6, \quad  \tilde \theta = y_7, \quad
 \frac{d \tilde \theta}{dz} = y_8,
      \label{matlab_eqns}
\end{eqnarray}
we obtain a set of 8 first order ODEs. The boundary conditions for this system are derived from equations (\ref{bcs_bottom}) and
(\ref{bcs_top}), noting that the no-slip bottom boundary implies $\omega_z=0$ at $z=0$ and the stress-free condition at the top implies
$ \partial \omega_z / \partial z =0$ at $z=1$:
\begin{eqnarray}
y_1=0, \quad y_2=0, \quad y_5=0, \quad y_8=0, \quad {\rm at} \quad z=0, \nonumber \\
y_1=0, \quad y_3=0, \quad y_6=0, \quad y_8=0, \quad {\rm at} \quad z=1.
\label{lin_bcs}
\end{eqnarray}
A small modification of the code used to solve this problem turns it into an eigenvalue code for the critical Rayleigh number when $F=0$. A boundary condition $y_2=1$ at $z=1$ is added, and the eigenvalue $Ra_F$ is minimised over $k_y$ to obtain
$Ra_{crit}$ and $k_{crit}$ appearing in Table \ref{table:cases3}.

Before examining the results from this model, we consider an asymptotic analysis of this system valid at large $Ta$.  

%
\subsection{Asymptotic analysis of the basic linear state model}

Here we assume $Ta$ is large, and that the solution stretches over three regions, a forced Ekman layer of thickness 
$O(Ta^{-1/4})$ near $z=1$, an interior region, and a normal Ekman layer of thickness $O(Ta^{-1/4})$ near $z=0$.
To be consistent with this choice, we need $H_s \sim H_f \sim O(Ta^{-1/4})$. 
From Appendix B equations (\ref{AppB_wsol_eq})  to  (\ref{AppB_aisol_eq}) if $F \sim O(Ta^{3/4})$ then $\tilde w \sim O(1)$ in the forced Ekman layer. As $\tilde w$ matches on to the interior layer and then to the bottom Ekman layer, $w \sim O(1)$ in both these regions too. In the two Ekman layers, $\partial / \partial z \sim O(Ta^{1/4})$ while it is $O(1)$ in the interior. To get the terms in 
(\ref{AppB_zcurlb_eq}) and (\ref{AppB_zdoublecurlb_eq}) to balance, we must have ${\tilde \omega}_z \sim O(Ta^{1/4})$ and hence $\tilde u \sim O(Ta^{1/4})$. We choose $Q \sim O(Ta^{1/4})$ and $H_s \sim O(Ta^{-1/4})$. We see below that this leads to  $\tilde \theta \sim O(1)$ in the forced Ekman layer, and it also matches onto the interior and the bottom Ekman layer so  $\tilde \theta \sim O(1)$
in those regions too. Now convection starts when $Ra_F \sim Ta^{2/3}$, and so with this scaling the term 
$k_y^2 Ra_F \tilde \theta$ in (\ref{lin_ord_omz}) is only $O(Ta^{2/3})$ while the other terms are much larger, $O(Ta)$. So this term is negligible with these scalings, so the Appendix B results for the velocities apply even when $\tilde \theta$ is $O(1)$. In the forced Ekman layer, $\tilde w$ is solved for explicitly in Appendix B, and since $Q \exp(-1/ H_s)$ is extremely small, in the forced Ekman layer (\ref{theta_eq_lin}) becomes
\begin{equation} 
\tilde w \left\{Q \exp \left(\frac{z-1}{H_s}\right) - 1\right\}=  \frac{d^2 \tilde \theta}{d z^2}-k_y^2 {\tilde \theta} .
 \label{asymp_temp_eq}
\end{equation}
We can use this to solve for $\tilde \theta$ explicitly in the forced Ekman layer. Not all the terms in (\ref{asymp_temp_eq}) are of the same order of magnitude, but as we apply derivative boundary conditions we retain them all here. 
Using the solution for ${\tilde w}$ in (\ref{AppB_wsol_eq}) and below of Appendix B, in the forced Ekman layer,
\begin{eqnarray} 
{\tilde \theta} = &t_0& \,\sinh (k_y z) + t_1 \cosh (k_y z) + \frac{a_0}{k_y^2} + t_2 \exp \left( \frac{z-1}{H_s} \right) 
+t_3 \exp \left\{(z-1) \left(\frac{1}{H_s} +Ta^{1/4}\frac{1+i}{\sqrt{2}} \right)\right\} \nonumber \\
+ &t_4&  \exp \left\{(z-1) \left(\frac{1}{H_s} +Ta^{1/4}\frac{1-i}{\sqrt{2}} \right)\right\}
+ t_5 \exp \left\{ (z-1) \left(\frac{1}{H_f} + \frac{1}{H_s} \right)\right\}
 \label{asymp_temp_top}
\end{eqnarray}
where
\begin{equation} 
t_2= a_0 Q H_s^2, \quad t_3 = a_1 Q\left( \frac{1}{H_s} + Ta^{1/4} \frac{1+i}{\sqrt{2}} \right)^{-2},  
\quad t_4= a_2 Q\left( \frac{1}{H_s} + Ta^{1/4} \frac{1-i}{\sqrt{2}} \right)^{-2},  
\quad t_5= a_3 Q\left( \frac{1}{H_s} + \frac{1}{H_f} \right)^{-2}.
\label{asymp_tdefs}
\end{equation}
In this expression for $\tilde \theta$ some terms have been neglected because they are small in the large $Ta$ limit compared with those retained. 
The terms that have cancelled are the exponential terms in (\ref{AppB_wsol_eq}) when multiplied by the $-1$ term in the basic state temperature gradient which are
$O(Ta^{1/4})$ smaller than the terms multiplied by $Q \exp ((z-1)/H_s)$, but note the $a_0$ part of $\tilde w$ multiplied by $-1$ must be retained.
Also for the exponential terms (but not for the $a_0$ term or the homogeneous part of the solution) the terms arising from
$-k_y^2 \tilde \theta$ are $O(Ta^{1/2})$ smaller than the terms arising from $d^2 {\tilde \theta} / dz^2$ and so are omitted. 

In the interior, the temperature equation is
\begin{equation}
- {\tilde w} = \frac{d^2 {\tilde \theta}}{dz^2} - k_y^2 {\tilde \theta} = - a_0,
\label{asymp_w}
\end{equation}
since ${\tilde w}$ is constant in the interior, because ${\tilde v}$ and hence $d {\tilde v} / dy$ is small there (no forcing in the interior). In the bottom Ekman layer the exponential terms in (\ref{AppB_wsolbot_eq}) only lead to $O(Ta^{-1/2})$ terms in $\tilde \theta$
there, which do not affect the bottom boundary condition at leading order. So from the boundary condition $d {\tilde \theta} / dz=0$ at $z=0$,  so $d {\tilde \theta }/ dz=0$ applies at the bottom of the interior layer as well as at $z=0$. So the temperature perturbation in the interior is
\begin{equation}
{\tilde \theta} = t_{int} \cosh (k_y z) + \frac{a_0}{k_y^2},
\label{asymp_tilde_theta}
\end{equation}
where $t_{int}$ is a constant determined by matching the forced Ekman layer temperature gradient to the interior temperature gradient. 
Matching the gradients gives 
\begin{equation}
   t_0 k_y \cosh k_y + t_1 k_y \sinh k_y =  k_y t_{int} \sinh k_y, 
\label{asymp_match_grad}   
\end{equation}
and matching the values gives
\begin{equation}
  t_0 \sinh k_y + t_1 \cosh k_y + \frac{a_0}{k_y^2} = t_{int} \cosh k_y + \frac{a_0}{k_y^2}, 
\label{asymp_match_values}
\end{equation}
so the solution is $t_0=0$ and $t_{int} = t_1$. 
Applying the boundary condition on $z=1$ that $d \tilde \theta / dz =0$, $t_1$ is determined by 
\begin{equation}
    - t_1 k_y \sinh k_y =  \frac{t_2}{H_s}  
    +t_3 \left( \frac{1}{H_s} + Ta^{1/4} \frac{1+i}{\sqrt{2}} \right) + 
    t_4  \left( \frac{1}{H_s} + Ta^{1/4} \frac{1-i}{\sqrt{2}} \right) +
    t_5 \left( \frac{1}{H_s} + \frac{1}{H_f} \right). 
\label{asymp_t1_def}
\end{equation}
The temperature perturbation is now determined everywhere. At $z=1$ and $z=0$
\begin{equation}
{\tilde \theta}(1)=t_1 \cosh k_y + \frac{a_0}{k_y^2}+t_2+t_3+t_4+t_5, \quad {\tilde \theta}(0) = t_{int} + \frac{a_0}{k_y^2}.
\label{asymp_temp_top_bot}
\end{equation}
The zonal flow ${\tilde u}$ can now be determined. In the interior there is thermal wind balance, 
\begin{equation}
Ta^{1/2} \frac{d {\tilde u} }{dz} = - k_y Ra_F {\tilde \theta},
\label{asymp_internal_ueq}
\end{equation}
so
\begin{equation}
{\tilde u} = - \frac{Ra_F t_{int}}{Ta^{1/2}} \sinh (k_y z) - \frac{a_0 Ra_F}{k_y Ta^{1/2}} z + U_c,
\label{asymp_internal_usol}
\end{equation}
and the constant $U_c$ is determined by the bottom Ekman layer matching. From equation (\ref{eq_noslip_wu_relation}) 
matching $w=a_0$ in the interior to $w$ at the top of the bottom Ekman boundary layer
gives 
\begin{equation}
 U_c =\frac{\sqrt{2} a_0}{k_y} Ta^{1/4},
\label{asymp_Uc_def}
\end{equation}
 so
\begin{equation}
 {\tilde u} =  - \frac{Ra_F t_{int}}{Ta^{1/2}} \sinh (k_y z) - \frac{a_0 Ra_F}{k_y Ta^{1/2}} z +
 \frac{\sqrt{2} a_0}{k_y} Ta^{1/4}
\label{asymp_internal_full_usol}
\end{equation}
in the interior region. 

In most of the simulations, $t_{int} = t_1$ is negative, while $a_0$ is positive, because $\tilde w$ is positive, see e.g. Fig.\,\ref{figs:fig2}. Usually, $|t_{int}|> a_0/k_y^2$ so from
equation (\ref{asymp_tilde_theta}) $\tilde \theta$ is negative, so that $\bar \theta$
is negative at $y=1$ and positive at $y=3$, as in Fig.\,\ref{figs:fig2}(e), which leads to the desired fall off of $\bar u$
with $z$ from the thermal wind equation. If there is no stable layer, 
$Q=0$, then the $t_{int}$ term is zero, and $t_{int} = t_1$ is positive as in Fig.\,\ref{figs:fig2}(d). However, in run A13, we noticed that 
$\tilde \theta$ was positive, despite that run having a very large $Q=600$ value, i.e there was a very strong stable layer. The explanation is that $H_s=0.00625$ was very small, so that the stable layer was very thin, and in equations (\ref{asymp_tdefs}), $1/H_s$ dominates $Ta^{1/4}$ and $1/H_f$,
and $Q H_s^2 \ll 1$. This makes $t_2$ to $t_5$ all small, so $t_1$ is small and hence $|t_{int}| < a_0/k_y^2$, making $\tilde \theta > 0$ so that 
$\bar u$ increases with  depth as in 
Fig.\,\ref{figs:fig2}(b). Not only must the stable layer be strong, it must also be thick enough so that the downwelling fluid in the forced Ekman layer brings positive $\bar \theta$ down into the interior at $y=3$.

\begin{figure*}[h]
\includegraphics[scale=0.95]{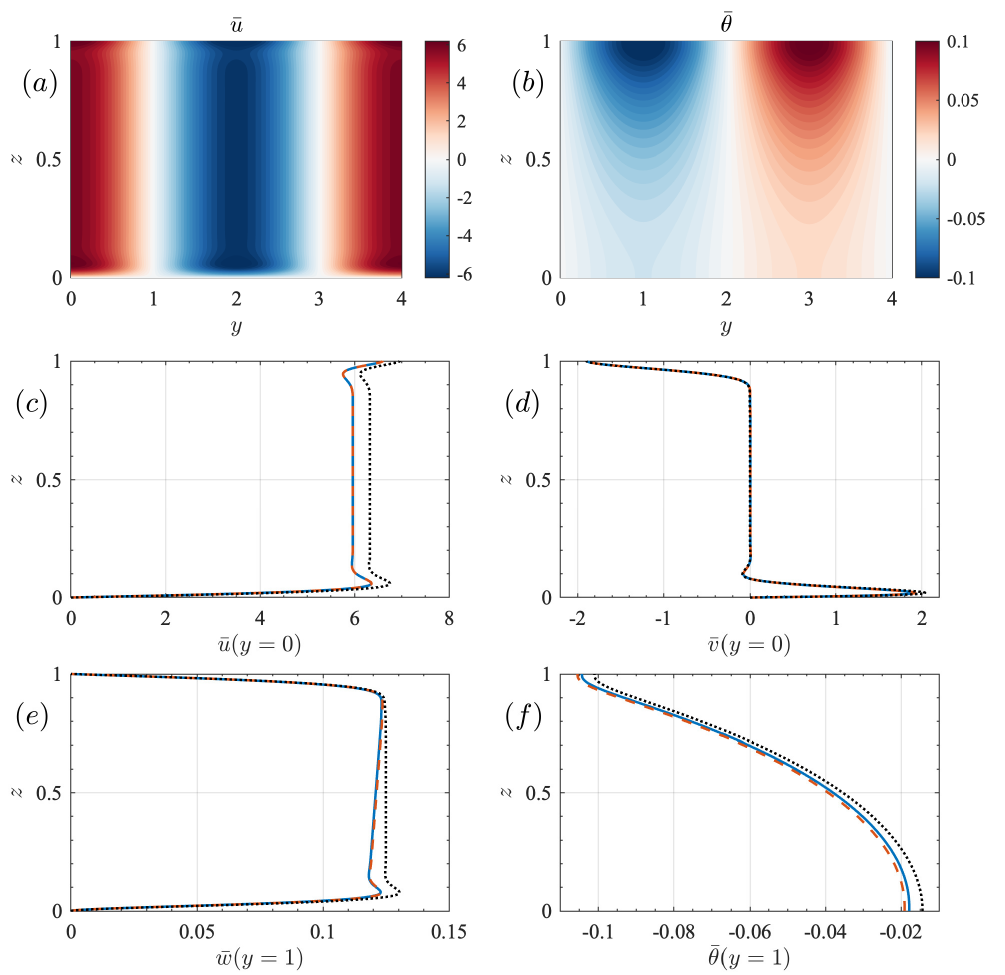} 
\caption{Comparison of the basic linear state model with simulations from run A14. (a) shows the mean zonal flow $\bar u(y,z)$.
(b) shows the mean temperature perturbation $\bar \theta$. Profiles from the nonlinear simulation (blue solid line), basic linear state model (orange dashed line) and asymptotic theory of the basic linear state model (black dotted line) are shown for $\bar u$ at $y=0$, $\bar v$ at $y=0$, $\bar w$ at $y=1$ and $\bar \theta$ at $y=1$ in (c), (d), (e) and (f), respectively.}
\label{figs:fig5}
\end{figure*}
\begin{figure*}
\includegraphics[scale=0.95]{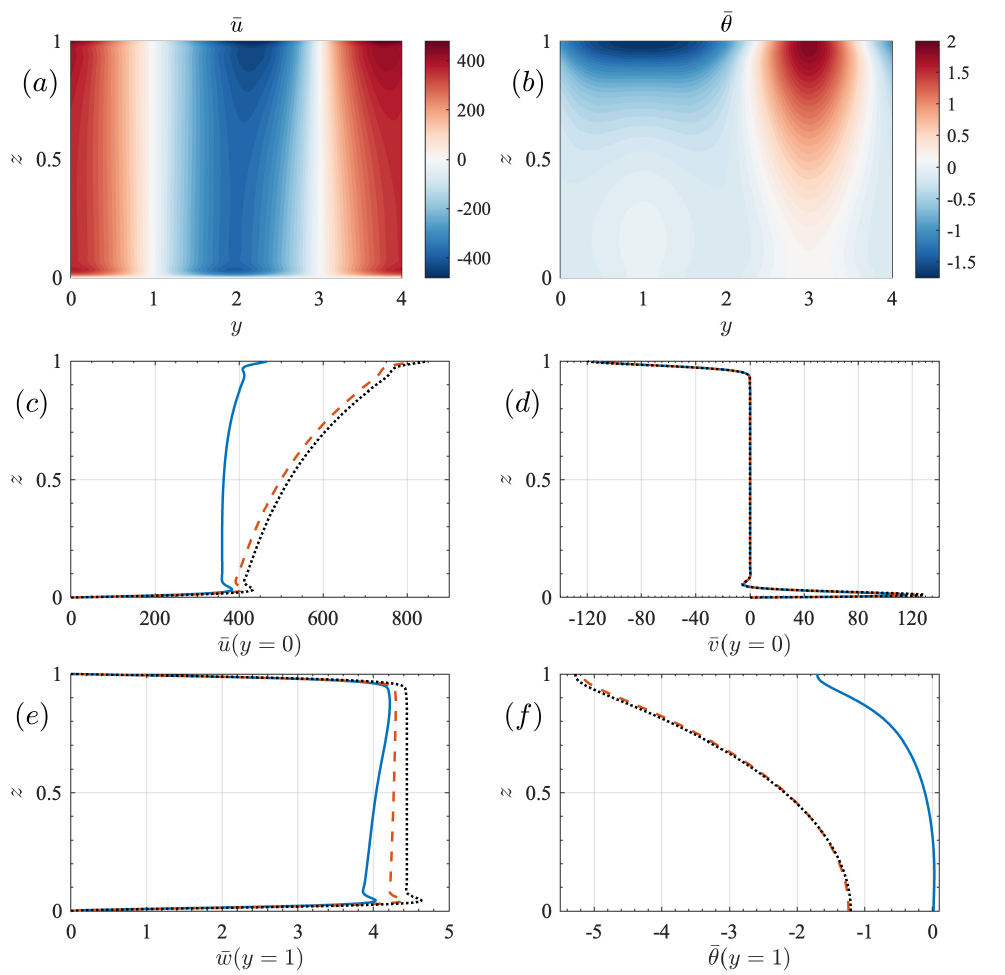} 
\caption{As for Fig.\,\ref{figs:fig5} but for simulation B2.}
\label{figs:fig6}
\end{figure*}

%
\subsection{Results for the basic linear state model}
%
%

\begin{figure*}
\includegraphics[scale=0.9]{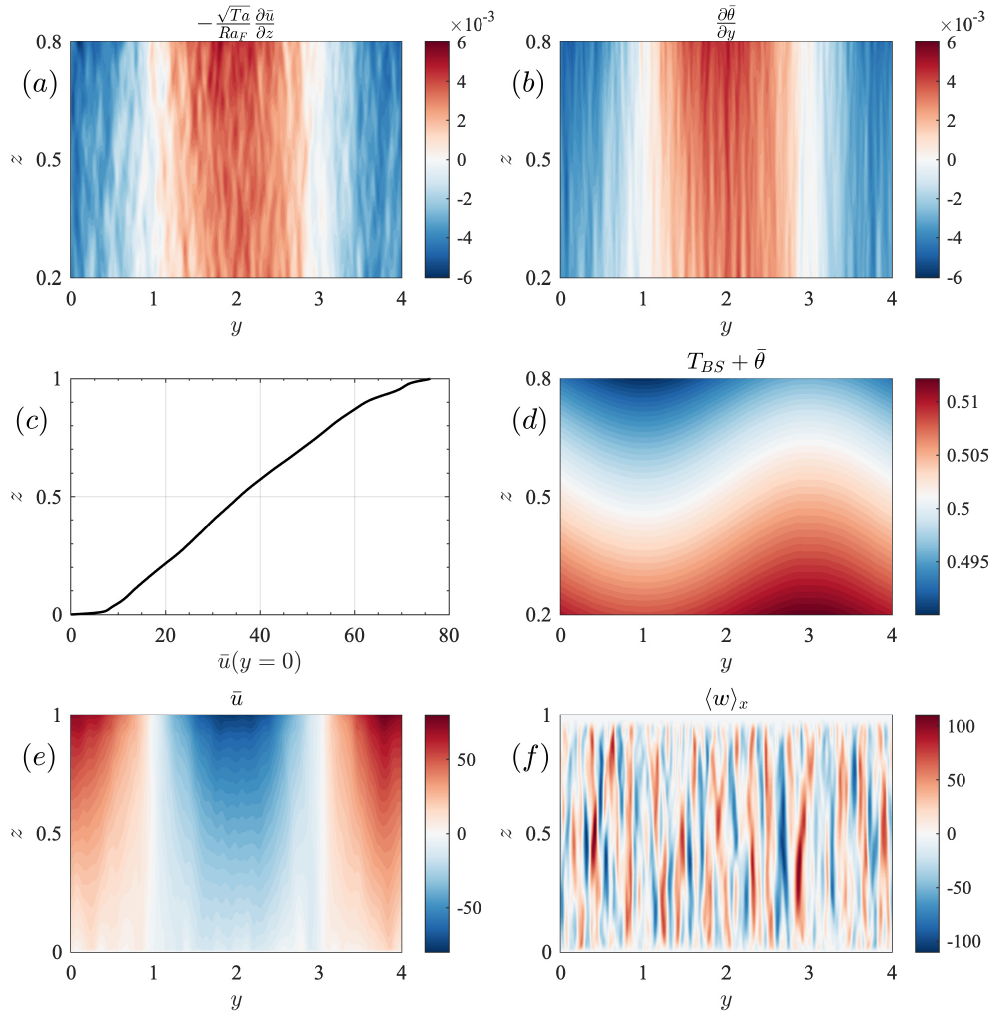} 
\caption{Results from simulation C1. (a) and (b) show the two terms in the thermal wind balance equation (\ref{thermal_wind_eq}). (c) shows the mean zonal flow ${\bar u}$ as a function $z$ at $y=0$. (d) shows the mean total temperature $T_{BS} + {\bar \theta}$ in the region $0.2<z<0.8$. 
(e) shows the mean zonal flow ${\bar u}$ as a function of $y$ and $z$ and (f) shows a snapshot of the $x$-averaged vertical velocity $\langle w \rangle_x$ }
\label{figs:fig7}
\end{figure*}

In Fig.\,\ref{figs:fig5} we compare the results of the basic linear state model with asymptotic results and with results from a full nonlinear simulation, run A14.
This run has no convection as the $Ra_F=0$, and the forcing $F$ is much less than $Ta^{3/4}$, so we expect the advection terms 
to be small in both the momentum and temperature equations. The agreement between the MATLAB results using (\ref{matlab_eqns}) and the full simulations is excellent, as expected. This also provides a useful check of the simulation code. The agreement between the simulation code and the asymptotic 
theory is not quite so good, as although $Ta$ is large, $Ta^{1/4}$ is not so big. Nevertheless, the form of the solutions given by the asymptotic theory is similar to that given by the MATLAB code, and as the behaviour of the asymptotic theory can be quickly examined 
as a function of the parameters, it is a useful tool in this problem. We also checked that at very large $Ta$, the results from the MATLAB code do approach those from the asymptotic theory very closely. 

In the basic state linear theory, the amplitude of the $\bar v$ and $\bar w$ components of velocity are simply proportional to $F$ and the form of the
velocity boundary layers is dependent only on $Ta$ and $H_f$. We only examined the case when the thickness of the Ekman layer, $Ta^{1/4}$, is similar to the 
forcing thickness $H_f$. The $y$-periodic forcing gives rise to periodic Ekman suction and pumping, generating a non-zero vertical velocity
in the interior, Fig.\,\ref{figs:fig5}(e), which becomes independent of $z$ as the interior is approached. Provided the stably stratified layer is strong, $Q \gg 1$, and is 
of comparable thickness to the forced Ekman layer, the vertical velocity in the interior leads to a $y$-periodic temperature at the bottom of the forced Ekman layer, as seen in Figs.\,\ref{figs:fig5}(b) and (f). This temperature anomaly is gradually reduced in the interior as $z$ decreases due to
thermal conduction in the $y$-direction. Note that $\bar v$ is only non-zero in the boundary layers, see Fig.\,\ref{figs:fig5}(d), in this basic state model.
When $Ra_F=0$ there is no thermal wind, so ${\bar u}$ is then constant in the interior, having been forced in the top Ekman layer. 

In Fig.\,\ref{figs:fig6}, we compare the basic state linear model with simulation B2, which has strong forcing $F$ and a non-zero $Ra_F$.
A number of important differences between the simulations and the basic state linear model emerge. Although $\bar v$ is still much smaller in the interior than in the boundary layers, it is not completely zero, because of the influence of the turbulent fluctuations due to convection and the shear providing non-zero Reynolds stresses. This means that $\bar w$ is no longer constant in the interior, it reduces slowly as the bottom boundary is approached. 
However, the main differences between the simulations and the linear model are in $\bar \theta$ and $\bar u$. The linear model predicts $\tilde \theta < -5$ in the top boundary layer, whereas the simulation gives $\tilde \theta \approx -1.7$ there. Inspection of the balance of terms in the temperature equation in the top boundary layer shows that in the linear case the balance is between the $d T_{BS}/dz$ term and the thermal conduction term, but in the simulation both the mean and fluctuating horizontal advection terms ${\bar v} \partial {\bar \theta} / \partial y$  and 
$\langle\partial / \partial y ( { v'} \theta' )\rangle _{xt}$ (which comes from the convection) are important in the top boundary layer, and their
effect is to substantially reduce the temperature $\bar \theta$, and hence the  horizontal temperature anomaly there. Not only is the magnitude of the horizontal
temperature anomaly at the top of the interior reduced below that expected from linear theory, it is further eroded by turbulent and mean flow advection in the interior, as can be seen in Fig.\,\ref{figs:fig6}(b). Since the thermal wind equation is still approximately valid in simulation
B2, the reduced horizontal temperature anomaly means that the zonal flow is much more constant than it is in the linear theory, see 
Fig.\,\ref{figs:fig6}(c). We have learnt from this that while the linear basic state model produces a zonal flow that gradually reduces with depth in the interior, as observed in Jupiter, the B2 simulation has only a rather small zonal flow difference between the top and bottom boundary layers, with almost all the variation in zonal flow occurring in the boundary layers themselves.

%
\subsection{Results for low $F$ and high $Ta$, model C1}
We now look at the results for a model with $Ta$ increased to $10^9$ as a large $Ta$ is more realistic for planetary interiors.
We focus on a strongly convecting case with $Ra_F=5 \times 10^8$ (run C1). The forcing $F=2 \times 10^5$, which is only $3.5\%$ of $Ta^{3/4}$, so the forcing is relatively weak. In Fig.\,\ref{figs:fig7}(a)
and (b) we see that thermal wind balance holds reasonably well even at this comparatively large $Ra_F$. Note that these plots still show some remnants of 
the columnar structure of the convection, despite being averaged over 2.28 diffusion times. In Fig.\,\ref{figs:fig7}(c)
we see the plot of $\bar u(z)$ at $y=0$ which we can compare with Fig.\,\ref{figs:fig2}(c). We see that the small value of $F$ leads to only a small jump in $\bar u$ across the boundary layers, so that unlike A9 (and many of the other A runs) most of the drop in $\bar u$ from its maximum near the surface to zero at the bottom occurs in the interior. Note that this occurs despite $\partial {\bar \theta} / \partial y$ being small, which is 
a consequence of $F$ being small. It is the large $Ra/\sqrt{Ta}$ factor in the thermal wind equation that compensates for the small $\partial {\bar \theta} / \partial y$ to give a significant $\bar u$ variation. In  
 Fig.\,\ref{figs:fig7}(d) the mean of the full temperature, $T_{BS}+ {\bar \theta}$, is shown. Note that in the interior the whole range of this quantity is within $T= 0.5 \pm 0.01$, showing that the temperature is approximately uniform in the interior due to the mixing arising from the convection. However, despite this mixing, the small horizontal temperature anomaly produced by the thin stable layer at the top is still visible in Fig.\,\ref{figs:fig7}(d). We attribute this to the strongly columnar nature of the convection, which mixes very efficiently in the $z$-direction but less efficiently in the horizontal direction. This mixing in the $z$-direction is why the temperature perturbation $\bar \theta$ almost exactly balances the basic state temperature, leaving only the small persistent horizontal variation seen in Fig.\,\ref{figs:fig7}(d). 
The mean of the zonal velocity ${\bar u(y,z)}$ can be seen in Fig.\,\ref{figs:fig7}(e).  It can be seen that ${\bar u}$ is approximately sinusoidal in $y$ at all values of $z$, so that although the zonal flow diminishes with depth, its $y$-variation is preserved deep into the interior. 
In Fig.\,\ref{figs:fig7}(f) 
a snapshot of the vertical velocity $w$ is shown, clearly showing the tall thin columns characteristic of rapidly rotating convection. If $\langle w\rangle _x$ is averaged over time, the amplitude of the velocity is reduced by cancellation as the thin columns move around in time, 
but the pattern remains similar. The columnar structure of the convection can also be seen in snapshots of ${\langle u(y,z,t)\rangle _x}$, and as the amplitude of the convective parts of $u$ and $w$ are similar, these columnar parts rather obscure the part which is more sinusoidal in $y$. Time averaging diminishes the amplitude of the time varying columns, which is why the smooth zonal flow is apparent in Fig.\,\ref{figs:fig7}(e).

\begin{figure*}
\includegraphics[scale=0.9]{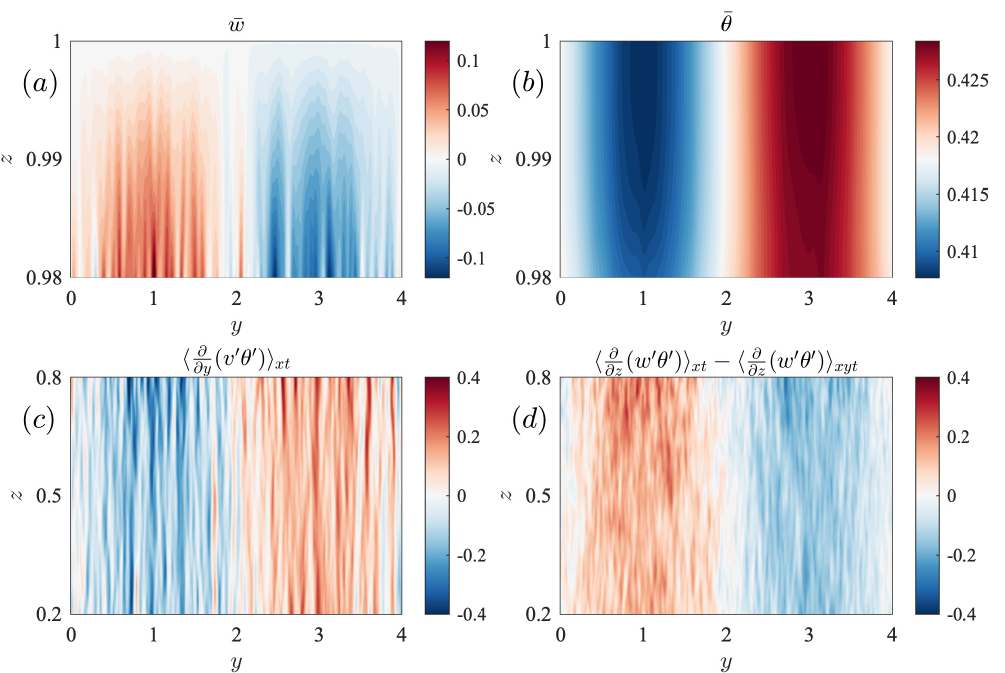} 
\caption{Results from simulation C1. (a) Mean vertical velocity $\bar w(y,z)$ and (b) mean temperature perturbation $\bar \theta(y,z)$ over the region $0.98<z<1$.
(c) Mean horizontal temperature advection from the fluctuating terms of equation (\ref{mean_temperature_eq}) and (d) mean vertical temperature advection. To make the sinusoidal part clearer in (d), the $y$ average of this term is subtracted off. Both (c) and (d) are for the interior region $0.2<z<0.8$. }
\label{figs:fig8}
\end{figure*}
 In order to investigate how the horizontal temperature variation, the horizontal temperature anomaly, is set up near the top boundary, we examine ${\bar w}$ and ${\bar \theta}$ close to the top boundary in Figs.\,\ref{figs:fig8}(a) and (b). In the region $0.98 < z <1$ (note the forcing layer thickness $H_f=0.01$ and the stable layer thickness $H_s=0.00815$) the columnar nature of $\bar w$ is apparent, but so is the sinusoidal component, positive near $y=1$ and negative near $y=3$. Note that this small, almost sinusoidal in $y$, time-averaged part of ${\bar w}$ is only visible because the amplitude of the convective part is reduced by the no-penetration boundary condition near the top and is further weakened by the time-averaging. If the length of the time-averaging was greatly extended, the columnar part would average to zero and only the forced sinusoidal part would be left
 (however, it is computationally expensive to do very long runs at this high $Ta$).  
 This small sinusoidal component of ${\bar w}$ acts on the strong stable stratification ($Q=460$) to produce the sinusoidal component of $\bar \theta$ shown in Fig.\,\ref{figs:fig8}(b). Note that the range of $\bar \theta$ in Fig.\,\ref{figs:fig8}(b) is not that large, as $\bar \theta$ is mixed by the vertical convection, but
 the sinusoidal part of $\bar \theta$ is an important component which leads to the thermal wind, and it persists almost to the base of the layer, note ${\bar \theta}_{vr} = 0.70$ from Table \ref{table:cases3}.
 The mean temperature is governed by equation (\ref{mean_temperature_eq}), and in Figs.\,\ref{figs:fig8}(c-d) we show the mean temperature advection in the interior derived from the fluctuating terms in angled brackets in (\ref{mean_temperature_eq}). Fig.\,\ref{figs:fig8}(c)  shows the horizontal advection $\langle \partial / \partial y (v' \theta') \rangle_{xt}$,
 while Fig.\,\ref{figs:fig8}(d)  shows the vertical advection with the $y$-averaged part subtracted off, $\langle \partial / \partial z (w' \theta') \rangle_{xt}$ - $\langle \partial / \partial z (w' \theta') \rangle_{xyt}$. The vertical horizontally averaged heat flux from the fluctuating terms is $\langle (w' \theta') \rangle_{xyt}$. Since the total horizontally averaged heat flux is $z$-independent, $\partial / \partial z \langle (w' \theta') \rangle_{xyt}$ is small, non-zero only due to conduction and advection by the large scale mean terms. However, subtracting off $\langle \partial / \partial z (w' \theta') \rangle_{xyt}$ does make the sinusoidal in $y$ part of $\langle \partial / \partial z (w' \theta') \rangle_{xt}$ clearer.  The small scale columnar structure is visible in these plots, but if the averaging were done over a longer timescale  the columnar part would average out and only the part sinusoidal in $y$, $\sim  \sin (\pi y / 2)$, would remain. As can be seen from Figs.\,\ref{figs:fig8}(c-d) these sinusoidal parts are almost equal in magnitude and opposite in sign.  From Fig.\,\ref{figs:fig8}(c) we see that the horizontal heat flux
 $\langle v' \theta' \rangle \sim f(z) \cos (\pi y /2)$ where $f(z)>0$, so the horizontal heat flux is trying to even out the horizontal temperature
 anomaly, and this is compensated for by the downward flux of temperature from the top boundary layer where the anomaly was set up. So the temperature anomaly must weaken as we go downward, but this horizontal flux is weak in runs C1 (and B3) compared to its value in other runs. This means the temperature anomaly penetrates deeper (larger ${\bar \theta}_{vr}$)
 than in other runs. We have not yet investigated why these runs with relatively low forcing and strong convection have a thermal anomaly which persists more deeply than in other runs. Possibly in other runs there is more shear induced turbulence which may be better at evening out the temperature. Another possibility is that the convective columns advect heat up and down the columns more efficiently than they spread it horizontally.

\subsection{Energy input into fluctuations}
So far we have assumed that the fluctuating parts of the variables have been due to convection. In many cases this is 
mainly correct, but it is possible that the zonal flow driven near the surface could lead to a flow that is unstable to $x$-dependent
perturbations. If this happens, then the energy source of the fluctuations could be the kinetic energy of the zonal flow rather than the convection. 
We did explore whether $x$-dependent fluctuations driven by the zonal flow could occur at zero Rayleigh number, i.e.  in the absence of convection.
We found no evidence that within the range of parameters we explored (with $F$ and $Ta$ in the range considered in Table \ref{table:cases3}) any
$x$-dependent instabilities occur, but they may well exist at much higher Reynolds numbers than we reached. Nevertheless, it is of interest to explore 
whether the fluctuations we did observe at finite $Ra_F$ are driven by convection or by Reynolds stresses arising from the driven flow.

Splitting the momentum equations into mean and fluctuating parts, as defined above, gives 
\begin{eqnarray}
\label{mean_fluct_eq}
    \frac{\partial {\bf u'}}{\partial t} &+& ({\bf \bar u} + {\bf u'}) \cdot \nabla ({\bf \bar u} + {\bf u'})
 + Ta^{1/2} Pr {\bf \hat z} \times ({\bf \bar u} + {\bf u'})
= - \nabla (\bar p + p') \\ \nonumber 
&+& Pr \nabla^2 ({\bf \bar u} + {\bf u'}) + Ra_F Pr (\bar \theta+\theta') {\bf \hat z}
+F \cos (k_y y) \exp \left(\frac{z-1}{H_f} \right) {\bf \hat x}
\end{eqnarray}
The mean equation, averaging over $x$ and $t$, is
\begin{equation}
    {\bf \bar u}\cdot \nabla {\bf \bar u} + \overline{({\bf u'} \cdot \nabla ){\bf u'}}
 + Ta^{1/2} Pr {\bf \hat z} \times {\bf \bar u} 
 = - \nabla \bar p  + Pr \nabla^2 {\bf \bar u}  + Ra_F Pr \bar \theta {\bf \hat z}
+F  \cos (k_y y) \exp \left(\frac{z-1}{H_f} \right){\bf \hat x} .
  \label{mean_only_eq}
\end{equation}
Subtracting (\ref{mean_only_eq}) from (\ref{mean_fluct_eq}) gives the fluctuating part equations
\begin{eqnarray}
 \label{fluct_only_eq}
    \frac{\partial {\bf u'}}{\partial t} &+&  ({\bf u'} \cdot \nabla ){\bf \bar u} +  ({\bf \bar u} \cdot  \nabla ){\bf u'}
   +\left( ({\bf u'} \cdot \nabla ) {\bf u'} - \overline{({\bf u'} \cdot \nabla ){\bf u'}} \right)
+ Ta^{1/2} Pr {\bf \hat z} \times {\bf u'} \\ \nonumber
&=& - \nabla p' + Pr \nabla^2 {\bf u'} + Ra_F Pr \theta' {\bf \hat z}.
\end{eqnarray}
We now take the dot product of this term with ${\bf u'}$
\begin{eqnarray}
\label{energy_eq}
    u'_i \frac{\partial {u'_i}}{\partial t} &+&  u'_i {\bar u}_j \frac{\partial}{\partial x_j} u'_i + 
     u'_i  u'_j \frac{\partial}{\partial x_j} {\bar u}_i     
     + u'_i \left[ u'_j \frac{\partial}{\partial x_j}  u'_i - 
     \left( \overline{({\bf u'} \cdot \nabla {\bf u'})} \right)_i     \right]  \\ \nonumber
   &=& - u'_i \frac{\partial}{\partial x_i} p' + u'_iPr \frac{\partial^2}{\partial x_j^2} u'_i + Ra_F Pr w' \theta' ,
 \end{eqnarray}
which can be written
\begin{eqnarray}
\label{energy2_eq}
    \left( \frac{\partial }{\partial t} + {\bar u}_j \frac{\partial}{\partial x_j} \right)
    \frac{u'{_i}^2}{2} &=& -
     u'_i  u'_j \frac{\partial}{\partial x_j} {\bar u}_i     
     - u'_i \left[ u'_j \frac{\partial}{\partial x_j}  u'_i -  
     \left( \overline{({\bf u'} \cdot \nabla {\bf u'})} \right)_i     \right] \\ \nonumber
     &-&  u'_i \frac{\partial}{\partial x_i} p' + u'_iPr \frac{\partial^2}{\partial x_j^2} u'_i + Ra_F Pr w' \theta'.
 \end{eqnarray}
We then integrate over the periodic volume and take a time average. On doing so, the term on the left vanishes, as does the pressure term on the right and we are left with a shear term $S$ (work done by Reynolds stresses), a nonlinear term $N$ (should be small when there
is a strong zonal flow), the viscous term $V$ (rate of working of the viscous terms) and the buoyancy term $B$, 
\begin{equation}
\label{eqn:KEfluc}
   V = S + N + B,
\end{equation}
where 
\begin{eqnarray}
    S&=& - \int_V \int_t u'_i u'_j \frac{\partial {\bar u}_i}{\partial x_j}\, dv dt, \quad N =  - \int_V \int_t u'_i u'_j \frac{\partial u'_i}{\partial x_j}\, dv dt, \\ \nonumber
    \quad V &=& - Pr \int_V \int_t u'_i \frac{\partial^2 u'_i}{\partial x_j^2}\, dv dt, \quad  B= Ra_F Pr \int_V \int_t  w' \theta'\, dv dt.
\end{eqnarray}
In B{\'e}nard convection the balance is between buoyancy and viscous forces, but here the shear term might be large. 
Note that in the $S$ term, the simulations suggest that the zonal flow, i.e. the $x$-component of ${\bar u}_i$ is the biggest term, so evaluating the integral of $u' ({\bf u'} \cdot \nabla ) {\bar u}$ would be sufficient.

%
\begin{figure}
    \includegraphics{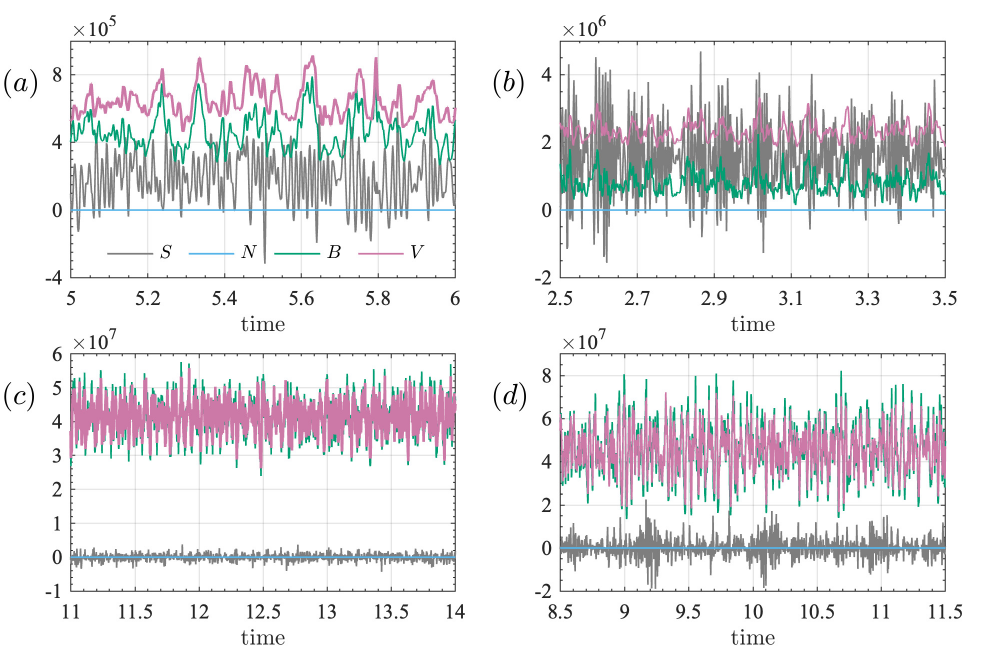}
    \caption{Terms in equation (\ref{eqn:KEfluc}) as a function of time for (a) Run A7, (b) Run A4, (c) Run B4 and (d) Run B5. The shear term $S$ is shown in grey, the nonlinear term $N$ in blue, the buoyancy term $B$ in dark green and the viscous term $V$ in pink.}
    \label{figs:fig9}
\end{figure}

In Fig.\,\ref{figs:fig9} the four terms $S$, $N$, $B$, and $V$ are shown as a function of time for four runs:
Fig.\,\ref{figs:fig9}(a) is for run A7, where the shear term $S$ is slightly lower than $B$ and $V$, but is nonetheless significant. Fig.\,\ref{figs:fig9}(b), run A4,  has the same $Ta$ and $Ra_F$, but the forcing is twice as big. Now the shear term $S$ is greater than the buoyancy term, so most of the fluctuation energy is derived from the shear of the imposed zonal flow.  The shear term fluctuates much more than $B$ and $V$ both when it is small and when it is large. 
Fig.\,\ref{figs:fig9}(c), run B4, has a higher $Ta$ and is strongly convecting, but it has a smaller forcing. Now there is almost a complete balance between buoyancy and viscous dissipation, but the shear term is just visible.  Fig.\,\ref{figs:fig9}(d), run B5, is for the same $Ta$ and $Ra_F$ as panel (c), but the forcing is twice as big. As expected this has enhanced the shear term, but the primary balance is still between buoyancy and viscous dissipation. The nonlinear term $N$ is negligible in all these runs. Run B5 exhibits periods in which the shear term $S$ is relatively weak and periods in which it is relatively strong. This suggests that it could be difficult to predict exactly when the shear term becomes significant, but it is clear that it always becomes important if the forcing is large enough.   

Note that even though run B4 has a small shear term as seen in Fig.\,\ref{figs:fig9}(c), it has a small $\bar\theta_{vr} = 0.18$ only. This seems surprising given that runs B3 and C1, where the shear is also small, had a much bigger value of $\bar\theta_{vr}$. However, the relatively weak forcing $F$ in both runs B3 and B4 mean that the thermal anomaly is relatively small where it is created in the top boundary layer. This means that even quite small shear induced fluctuations, as in run B4, are enough to reduce $\bar\theta_{vr}$ to a small value. Only when the shear induced fluctuations are really small, as in run B3, do values of $\bar\theta_{vr}$ near one occur. We also note that the largest shear case, run A4, has a moderate $\bar\theta_{vr} = 0.45$. However, in this run the forcing is so large that the thermal wind equation is no longer satisfied well, and the top boundary layer has broadened out, so this run in a different regime, far from that of the giant planets.

\subsection{Results with large forcing and varying Rayleigh number}

%
\begin{figure}
    \includegraphics[scale=0.95]{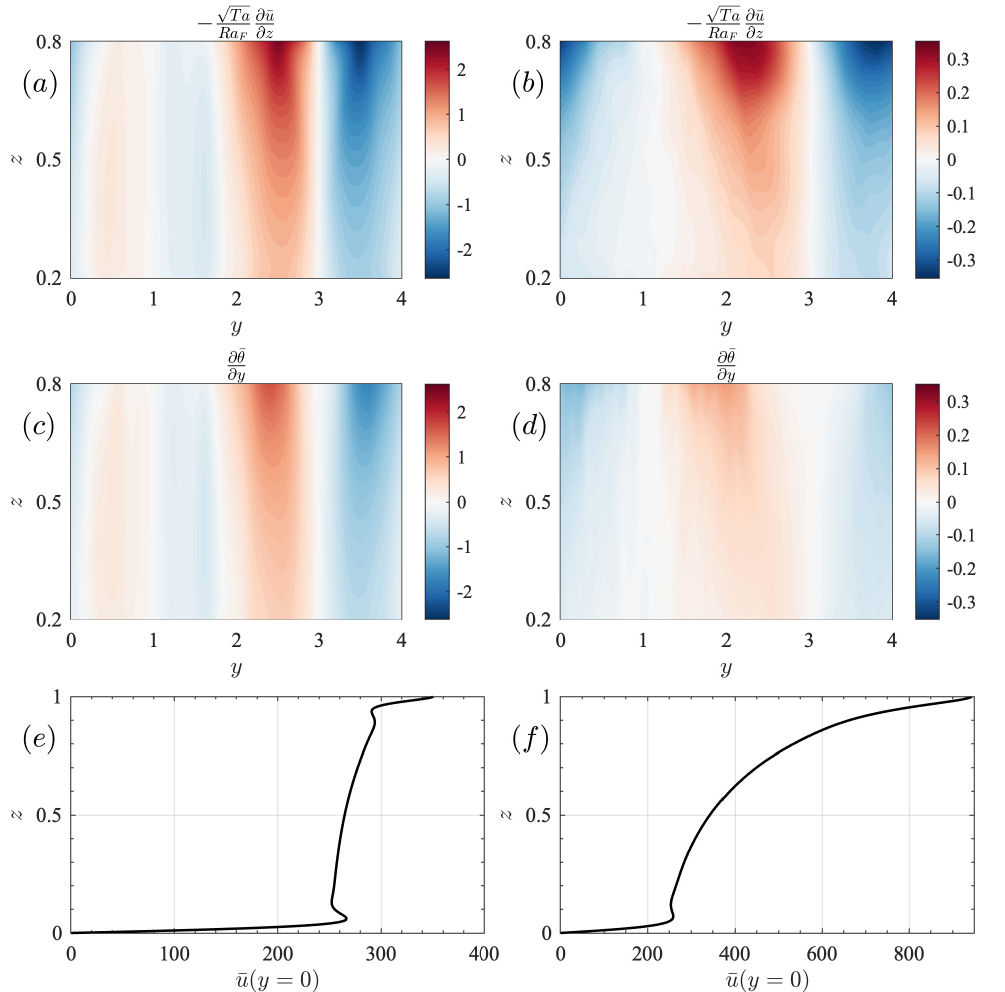}
    \caption{Thermal wind balance and $\bar u(y=0)$ for run A2 (left column) and run A6 (right column). Panels (a) and (b) show the right-hand-side of (\ref{thermal_wind_eq}) and (c) and (d) show the left-hand-side of (\ref{thermal_wind_eq}). If thermal wind balance (\ref{thermal_wind_eq}) holds exactly (a) and (c) should be identical  
    and so should (b) and (d). (e) and (f) show the mean zonal flow ${\bar u}(z)$ at $y=0$.}
    \label{figs:fig10}
\end{figure}

%
\begin{figure}
    \includegraphics[scale=0.95]{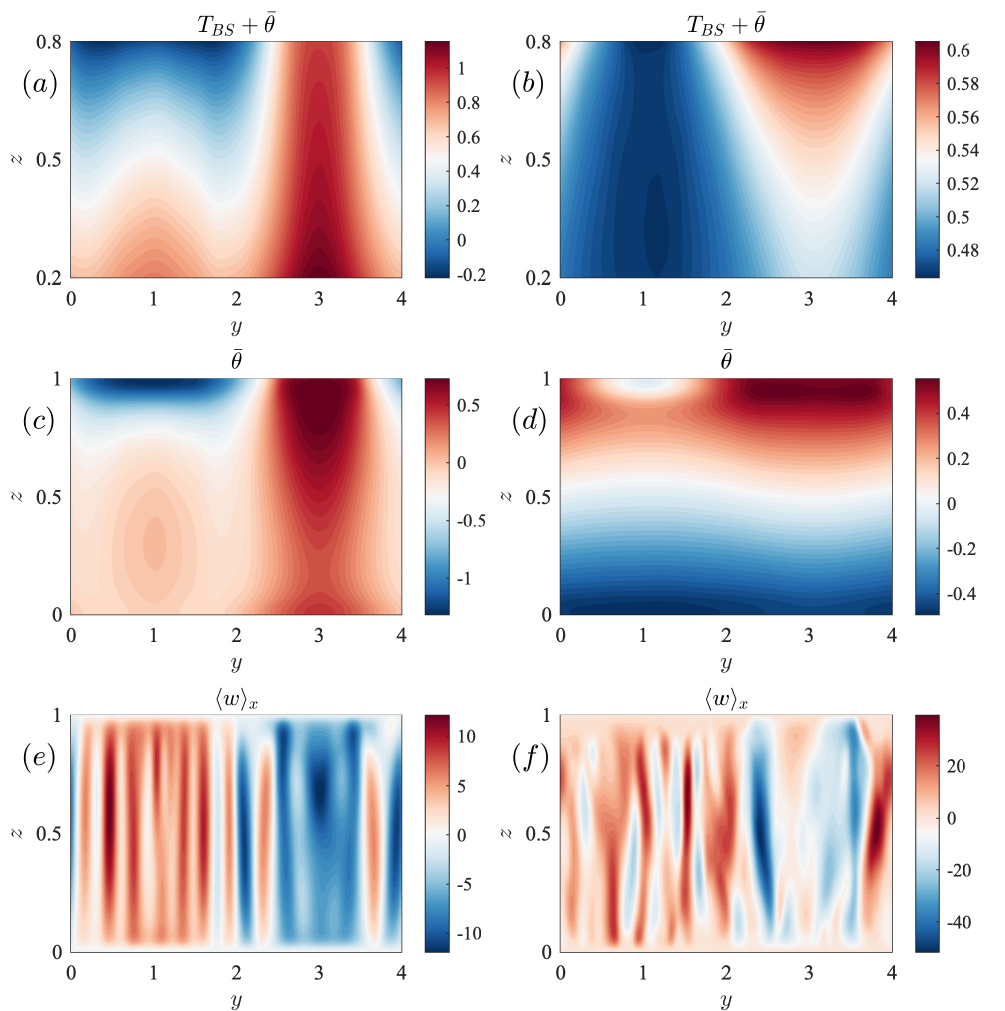}
    \caption{Further comparisons between runs A2 (left column) and A6 (right column).
    (a) and (b) show the mean total temperature $T_{BS}+{\bar \theta}$ in the region $0.2<z<0.8$.
    (c) and (d) show the mean temperature perturbation ${\bar \theta}$ over the whole layer. (e) and (f) show the $x$-averaged vertical velocity at a snapshot in time, $\langle w\rangle_x$. (e) still shows a columnar structure, but in (f) the columnar structure has almost broken up.}
    \label{figs:fig11}
\end{figure}

A sequence of runs with $Ta=10^7$ and $F\approx3.56 \times Ta^{3/4}$, i.e a strong forcing term, with the $Ra_F$ varying from 0 to
$Ra_F=Ta=10^7$ was performed. Comparing the $Ra_F=0$ case with the basic state linear model, we find that while the jumps in $\bar u$ across the boundary layers are similar to those predicted by linear theory, the temperature perturbations are  very different, as was the case in Fig.\,\ref{figs:fig6}(f) for run B2. 
When $Ra_F$ is non-zero, thermal wind balance can apply. At the smaller values of $Ra_F$, e.g. run A2, the thermal wind equation 
(\ref{thermal_wind_eq}) is satisfied reasonably well (see Figs.\,\ref{figs:fig10}(a) and (c)) but at the largest value of $Ra_F$, run A6, the inertial terms in (\ref{full_thermal_wind_eq}) are significant and thermal wind balance is poor (see Figs.\,\ref{figs:fig10}(b) and (d)). This means that the
zonal flow profile changes shape with increasing $Ra_F$, see Figs.\,\ref{figs:fig10}(e) and (f). At smaller $Ra_F$, Fig.\,\ref{figs:fig10}(e),
the jump in zonal flow $\bar u$ across the boundary layers is quite large, because $F$ is large and so the magnitude of $\bar u$ is large, but the change in $\bar u$ across the interior is relatively small. However, at this $Ra_F$ the temperature perturbation only falls off slowly with depth so the zonal flow gradient in $z$ is fairly constant. At larger $Ra_F$, Fig.\,\ref{figs:fig10}(f), the drop across the interior is larger because the factor $Ra_F/\sqrt{Ta}$ is larger, but the temperature perturbation gets washed out at depth (as explored in the next paragraph), so the zonal flow diminishes rapidly at the top but more slowly as we move into the interior. The net effect is that the top boundary layer is hardly 
visible in Fig.\,\ref{figs:fig10}(f). The profile is quite Jupiter-like, but the Rossby number of the convective part of the flow is much larger than
in Jupiter, so this may be just a coincidence. 

The total temperature, $T_{BS} + {\bar \theta}$, in runs A2 and A6 are shown in Figs.\,\ref{figs:fig11}(a) and (b) in the interior $0.2 < z < 0.8$.
Figs. (c) and (d) show ${\bar \theta}$ over the whole range $0<z<1$. At the top of the layer, panel (c) shows the temperature profile is almost sinusoidal in the top
boundary, though the cool (blue) patch is slightly broader than the hot (red) patch. The higher $Ra_F$ case, run A6, is much less sinusoidal in the top boundary layer, as the strong convection is affecting this region, as the nonlinear advection terms omitted in the basic linear state model are now very large. 
Fig.\,\ref{figs:fig11}(a) shows that in the lower $Ra_F$ case, A2, the sinusoidal part of the temperature perturbation extends
right to the bottom of the layer, which is why the zonal flow gradient only weakens by a relatively small amount in this case. However,
in Fig.\,\ref{figs:fig11}(b) the total temperature has been more thoroughly mixed, so the horizontal temperature anomaly driving the thermal wind
is much weaker at the bottom, so the zonal flow has not penetrated as deeply in this case. Also, in run A6 the thermal wind balance is not strong as the convective Rossby number is no longer small. 

Figs.\,\ref{figs:fig11}(e) and (f), which show snapshots of the $x$ averaged vertical velocity $\langle w\rangle _x$ give further evidence of why the horizontal temperature
gradient at the top does not penetrate down so far in run A6. In run A2, the convection columns still extend to the base of the layer, so they are able
to transmit the horizontal temperature perturbation almost to the bottom of the layer, but in run A6 the columns break up before they reach the bottom,
so the horizontal gradient at the top is spread out by horizontal advection before it reaches the bottom of the layer.

\section{Discussion and Conclusions}

A model of a rotating convecting layer with zonal flow forcing in a thin region near the top has been examined. This is motivated by the strong zonal flows on giant planets, which may be forced by the Reynolds stresses associated with the strongly turbulent flow in their surface layers. We found that a stably stratified layer near the surface was necessary to reproduce the zonal wind depth profiles revealed by the gravity measurements of Jupiter. A suite of fully three-dimensional 
numerical simulations has been performed to investigate the range of possible behaviour in the convecting layer. As expected, at large $Ta$ strong convection takes the form of time-dependent columnar motions with a short horizontal wavelength. The type of solution found
and the depth of the zonal flow depend on the parameters chosen, particularly the strength of the forcing and the Rayleigh number. Most solutions were found to be in thermal wind balance to a good approximation. When the forcing is relatively weak, small compared to $Ta^{3/4}$, and the convection was strong, simulations showed  that the horizontal temperature anomaly generated in the forced stably stratified layer extended deep into the interior. Then the
thermal wind balance means that the jets extend deep into the interior, as occurs on Jupiter. The solutions contained both a large-scale mean circulation with a zonal flow, together with a small-scale fluctuating component. The fluctuating component was mainly forced by thermal convection when the forcing was weak, but in more strongly forced cases, the Reynolds stresses arising from the shear of the strong zonal flow drive the fluctuations. With stronger forcing, the jets do not penetrate as deeply into the interior, but the penetration depth increases as $Ra$ increases. This suggests that whether the fluctuations are driven by convection or by shear can have an effect on how deep the flows go.

When there is no forcing in the top boundary layer, our problem reduces to a standard Rayleigh-B\'enard convection problem with a thin stably stratified 
layer at the top, and slightly unusual boundary conditions. This problem has a well-defined stationary basic state with a critical Rayleigh number $Ra_{crit}$ at which the solution becomes three-dimensional with a short horizontal wavelength $O(Ta^{-1/6})$. If $F\ne 0$ there is no stationary basic state, but instead a steady two-dimensional flow varying only in $y$ and $z$. If $F/Ta^{3/4}$ is small,  this state can be found approximately using the basic linear state model developed in section \ref{sec:linear}. As $Ra_F$ is increased, this 2D state becomes unstable to three-dimensional disturbances with the unstable modes having a short wavelength in the $x$ direction. The exact nature of this instability has not been fully investigated, but with the run A14 values of $Ta$, $F$ and $Q$, there is evidence that the bifurcation to $x$-dependence can be subcritical. Both an $x$-independent solution and an $x$-dependent solution were found at the same value of $Ra_F$. In these circumstances the finite amplitude $x$-dependent solution is never both small amplitude and stable, so nonlinear terms (ignored in the basic state linear model)
are important, so unfortunately we have not yet found convecting $x$-dependent solutions where the basic state linear model works well. If such regions of parameter space exist, they are likely to be sparse.

It is the forcing that gives this problem its distinctive character. The forcing leads to downwellings and upwellings in the forced boundary layer which together with the stable stratification at the top give rise to a horizontal temperature anomaly. In our  model, the forced boundary layer is viscous, with the turbulent Reynolds stresses being represented by the forcing 
$F \cos (k_y y) \exp((z-1)/H_f) \hat{\mathbf{x}}$, and when the buoyancy and nonlinear advection can be ignored in this boundary layer it has the character of a forced Ekman layer as described in Appendix B. The top boundary layer in giant planets will be controlled by Reynolds stresses, with molecular diffusion being unimportant. However, these stresses will have an eddy diffusion part and a zonal forcing part, and the observation of Ferrel-like cells \cite{Duer21} suggests that our crude model does capture these upwellings and downwellings which lead to the formation of a horizontal temperature anomaly.

Provided the Rossby number of both the mean flow and the turbulent flow is small, thermal wind balance holds reasonably well,
as can be seen from the small values of $\epsilon_{TW}$ in Table \ref{table:cases3}. Then the horizontal temperature anomaly set up in the forced boundary layer leads to a steady decay of the zonal winds as we go into the interior. In some runs, e.g. C1 and B3 where
$F/Ta^{3/4}$ is small and the convection is active (higher $Ra_F$) the horizontal temperature
anomaly persists throughout the interior, but in other runs  where the forcing is stronger and the convection is weaker, such as run B2, the horizontal temperature anomaly decays quite quickly with depth, ${\bar \theta}_{vr}$ being only 0.088 in this case. Increasing the Rayleigh number at fixed forcing, as in runs A1-6, also increases the depth to which the horizontal temperature anomaly reaches down. In most cases this leads directly to a deeper zonal flow because the thermal wind relation holds, but at high enough $Ra_F$ the thermal wind relation breaks down and the top boundary layer merges into the interior, see 
Figs.\,(\ref{figs:fig10},\ref{figs:fig11}). When $F/Ta^{3/4}$ is $O(1)$ and $Ra_F$ is small the zonal flow itself can drive the fluctuations, whereas when the forcing is weaker and the convection stronger the convection drives the fluctuations. It seems that when the fluctuations are driven by the convection the zonal flows penetrate deeper than when they are driven by the shearing motion from the zonal flow itself. This may be why the zonal flows seem to penetrate deeper than the Great Red Spot \cite{Parisi21}; possibly the turbulence there is more affected by the shear, so reducing the penetration depth. A possible reason the convection leads to deeper flows appears to be that the tall thin convective columns are good at transporting the horizontal temperature anomaly vertically
without spreading it out laterally. 

Our model has similarities to that of \cite{Christensen20}, which was later refined in \cite{CW24}. Their model  also has the zonal flow driven  by an explicitly  imposed force in the azimuthal direction located near the top, though our forcing was more concentrated at the top than theirs. The main difference is that in 
\citep{Christensen20, CW24} the stable layer was at the bottom of the layer rather than at the top. Nevertheless, in both models the forcing drives a meridional circulation which interacts with the stable layer to set up a horizontal temperature gradient in the $y$-direction (latitudinal direction). Since thermal wind balance holds in both models, this breaks the geostrophy of the zonal winds, allowing them to be quenched at depth. However, to get the horizontal temperature gradient in the right place, \citep{Christensen20, CW24} require the stable layer to be located at only $\sim 2000$\,km below the surface, much higher than Jupiter models suggest is likely. In our model, with a stable layer near the surface layers, the physical origin of the stable layer is clear, but the challenge is to get the horizontal temperature gradient to go deep enough, though our models suggest this is possible if the turbulence is dominated by strongly anisotropic rotating columnar convection. Another issue
common to both models is that while they have the zonal flows stronger than the convection, estimates of the convective velocity in Jupiter are less than 0.1 m/s
    (see e.g. Supplementary Table 1 in \cite{Hori23}) so the true ratio of $Ro_{\bar u}$ to $Ro_{vw}$ is greater than the values in Table \ref{table:cases3}. The convection needs to be strongly supercritical, but the convective velocities should nevertheless be small compared to the zonal wind. In giant planets this is achieved by very small viscosity, but this is a difficult limit for simulations, so some compromise is inevitable.

Our model can also be compared with that of Heimpel et al. \cite{Heimpel22}, who studied convection in a thin rotating shell with a stably stratified layer near the surface, with a view to seeing if the anticyclonic vortices and the zonal flows could be reproduced. The main dynamical difference between their work and ours
(besides the different geometry) is that they did not put in an explicit forcing, but relied on the Reynolds stresses produced by their convection all over the layer to drive the jets, i.e. they had deep, rather than shallow, forcing. They were successful in finding anticyclonic vortices, and their model also had alternating zonal jets. However, because they had no forcing at the top, they did not have a meridional circulation, so that the horizontal temperature
gradients found in our work and the \citep{Christensen20, CW24} models did not occur. Their jets were therefore very geostrophic, with a similar flow speed
at the top and the bottom of the layer, which is not compatible with the gravity data. Had they been able to extend their model to include the low density near surface regions,
where convective velocities are faster and so Reynolds stresses larger, they might have found meridional Ferrel-like cells, but despite using state-of-the-art numerical methods they had to cut their model off at 350\,km below the surface to properly resolve the computation.

In our model, the zonal flow is brought to rest at the bottom by a rigid boundary. In a giant planet it is the magnetic field that ensures the zonal flow is small deep down. The convection in the dynamo region does lead to zonal flows there, but these are of the order of cm/s rather than the tens of m/s observed higher up. Our model shows that it is the depth at which lateral diffusion (turbulent or laminar) smooths out the horizontal temperature anomaly that determines the depth of the fast (O(m/s)) zonal flows, not the depth of the field itself. At depths where the horizontal temperature anomaly is eliminated the flow is geostrophic, $z$-independent, and therefore constrained to very low values by the magnetic field. It is only where the horizontal temperature anomaly is significant that the winds can begin to build up to their large surface values.  

There are many aspects of our model which could be added to make it a more realistic model for a giant planet. One is to add a magnetic field at the bottom of the layer with a variable conductivity as in giant planets. As we approach the surface, the electrical current strengths must fall off rapidly, but we do not know how small they can be and still affect the zonal flow, so we do not know how important field is at 3500\,km below Jupiter's surface. Another natural development would be to make the problem anelastic. There is a large density variation between the uppermost regions where the observed turbulent forcing of the zonal flow occurs and the deep zonal flow thousands of km below. Is it really possible for this low density `tail' region to wag the dog of the deep zonal flow? The compressible thermal wind equation suggests this might be possible, but anelastic models would shed light on the conditions needed for this to happen. Finally, our model takes no account of the spherical geometry of the planetary problem. We did look briefly at models with a rotation vector tilted in the $yz$-plane, and they seemed to behave in a broadly similar manner with convection enhancing transport parallel to the rotation axis and restricting it perpendicular to that axis, but a more thorough investigation of geometrical effects is needed.

\begin{acknowledgments}
We thank the anonymous referees for their thoughtful reviews which improved the paper. LKC also gratefully acknowledges support from the STFC (ST/X001083/1 and ST/Y002296/1). This work has made use of the Hamilton HPC Service of Durham University and the DiRAC Data Intensive service (DIaL3) at the University of Leicester, managed by the University of Leicester Research Computing Service on behalf of the STFC DiRAC HPC Facility (www.dirac.ac.uk). The DiRAC service at Leicester was funded by BEIS, UKRI and STFC capital funding and STFC operations grants. DiRAC is part of the UKRI Digital Research Infrastructure. The authors would also like to thank the Isaac Newton Institute for Mathematical Sciences, Cambridge, for support and hospitality during the programme DYT2 where this paper was conceived. DYT2 was supported by EPSRC grant EP/R014604/1. For the purpose of open access, the author(s) has applied a Creative Commons Attribution (CC BY) licence to any Author Accepted Manuscript version arising.
\end{acknowledgments}

\section*{Data Availability}
The data that support the findings of this article are openly available \cite{Data}.

\appendix

\section{Zonal flow model of Kaspi et al. \cite{Kaspi23}}

The zonal flow model of Kaspi et al. \cite{Kaspi23} is derived from the observed zonal flow \cite{TOLL17} as a function of planetocentric latitude $\lambda'$
(the perijove 3 data set \cite{TOLL17} converted to planetocentric latitude was used here) to give $U^{obs}(\lambda^{\prime})$. A velocity profile projected down
parallel to the rotation axis as a function of $r$ and $\lambda$, $r$ being the distance from the planet's centre and $\lambda$ being the point's latitude
is constructed using
\begin{equation}
    U^{proj}(r,\lambda) = U^{obs}(r, \lambda^{\prime}), \quad \cos \lambda^{\prime} = \frac {r \cos \lambda}{R_J}  
\label{eq:Kaspi1}    
\end{equation}
where $R_J$ is the mean radius of Jupiter, 69946 km. This projected velocity profile is now multiplied by a radial decay function which brings
the projected velocity profile down towards zero at depth. The decay function that best fitted the gravity data was
\begin{equation}
  Q(r) = (1-\alpha) \exp\left( \frac{r -R_J}{H} \right) + \alpha \left[ \frac{ \tanh\left( \frac{H +r -R_J}{\Delta H}\right) +1}{\tanh\left(\frac{H}{\Delta H}\right) +1} \right] ,\quad U(r,\lambda) = U^{proj}(r, \lambda) Q(r),
\label{eq:Kaspi2}
\end{equation}
and $U(r,\lambda)$ is the accepted zonal flow profile in Jupiter's interior and it fits the gravity data rather well. The parameters used here are
$\alpha=0.68$, $H=2101$\,km and $\Delta H=824$\,km \cite{Kaspi23}. Since $Q(R_J)=1$ the
profile agrees with the observed surface profile. Fig.\,\ref{figs:fig1}(a) is a cross-section of this flow profile in the $s-z$ plane, where $s= r \cos \lambda$ and
$z= r \sin \lambda$. The dashed curve is the radius at $r= R_J - R_{depth}$ where $R_{depth}$ = 3500\,km and the jets have almost completely decayed at that depth. The two jets at
latitudes $21^\circ$\,N and $18^\circ$\,S contribute most to the signal. Fig.\,\ref{figs:fig1}(b) focuses on the jet at $21^\circ$\,N  which has $s_j=0.9335 R_J$.
Since the jet is primarily aligned with the rotation axis (see Fig.\,\ref{figs:fig1}(a), 
using equations (\ref{eq:Kaspi1}) and (\ref{eq:Kaspi2}) we see how the zonal velocity drops off with $z$ keeping $s$ constant at $s=s_j$, where $s_j^2 + z^2 = r^2$.
The depth of the jet in the $z$-direction is given by 
\begin{equation}
    z_d = R_J  \sin 21^\circ - \left[ (R_J-R_{depth})^2 - R_J^2 \cos^2 21^\circ  \right]^{1/2} = 12780 \, {\rm km.}
\end{equation}
The scaled $z$ plotted in Fig.\,\ref{figs:fig1}(b) is the physical $z$ in km divided by 12780, and the velocity is in metres/sec.

\section{ Boundary layer theory}

\subsection{Forced Ekman layer at the top near $z=1$}

The forcing is in a thin layer near the top boundary, $z=1$. The top boundary has no penetration and is stress-free. In that thin layer, buoyancy is small in most of our simulations. Although the forcing is presumed to be from small-scale turbulence acting through Reynolds stresses, the inertial terms arising in the simulation are often smaller than the viscous, Coriolis, and forcing terms in the top boundary layer. We can therefore 
get useful information from the $x$ and $t$ averaged equations by omitting the inertial and buoyancy terms. After solving this linearised system, we estimate the buoyancy term and the inertial terms to see when their neglect can be justified.  

We assume that the forcing is in the $x$-direction and has the form $F \cos (k_y y) \exp [(z-1)/H_f]$. The mean momentum equation in the layer has the form
\begin{equation}
    \nabla {\bar p} + Ta^{1/2} Pr {\bf \hat z \times  \bar u}  + \underbrace{ 
    {\bf \bar u}\cdot \nabla {\bf \bar u} +
    \left\langle{\bf u'}\cdot \nabla {\bf u'}\right\rangle }_{\text{inertial}}  = 
    Pr \nabla^2 {\bf \bar u} + F \cos (k_y y) \exp \left( \frac{z-1}{H_f} \right){\bf \hat x}  + \underbrace{ Ra_F Pr \bar \theta {\hat z}}_{\text{buoyancy}}
    \label{AppB_mom_eq}
\end{equation}
The $z$-component of the curl of (\ref{AppB_mom_eq}) is 
\begin{equation}    \label{AppB_zcurl_eq}
   -Ta^{1/2} Pr \frac{\partial {\bar w}}{\partial z}
   -\underbrace {\frac{\partial}{\partial y} \left( {\bar v} \frac{\partial {\bar u}}{\partial y} + {\bar w} \frac{\partial {\bar u}}{\partial z} \right) - \frac{\partial}{\partial y} \left\langle { v'} \frac{\partial {u'}}{\partial y} + {w'} \frac{\partial {u'}}{\partial z} \right\rangle  }_{\text{inertial}} 
   = Pr \frac{\partial^2 {\bar \omega_z}}{\partial z^2} + F k_y \sin(k_y y) \exp \left( \frac{z-1}{H_f} \right), 
\end{equation}
where
\begin{equation}
    {\bar \omega_z} = {\bf \hat z \cdot \nabla \times {\bar u}} = \frac{\partial {\bar v}}{\partial x} -\frac{\partial {\bar u}}{\partial y} = -\frac{\partial {\bar u}}{\partial y}
     \label{AppB_omzdef_eq}
\end{equation}
since the mean velocity is  independent of $x$. Note that $\nabla^2 {\bar \omega_z} \approx \partial^2 {\bar \omega_z}/ \partial z^2$
because the layer is thin. The $z$-component of the double curl of (\ref{AppB_mom_eq}) gives
\begin{eqnarray}    \label{AppB_zdoublecurl_eq}
    Ta^{1/2} Pr \frac{\partial {\bar \omega_z}}{\partial z} 
     &+&  \underbrace {\frac{\partial}{\partial y} \left[
    \frac{\partial}{\partial y}\left( {\bar v} \frac{\partial {\bar w}}{\partial y} + {\bar w} \frac{\partial {\bar w}}{\partial z} \right)
    -\frac{\partial}{\partial z}\left( {\bar v} \frac{\partial {\bar v}}{\partial y} + {\bar w} \frac{\partial {\bar v}}{\partial z} \right)\right]}_{\text{inertial}} \\ 
    &+& \underbrace {\frac{\partial}{\partial y} \left\langle
    \frac{\partial}{\partial y}\left( {v'} \frac{\partial {w'}}{\partial y} + {\bar w} \frac{\partial {w'}}{\partial z} \right)
    -\frac{\partial}{\partial z}\left( {v'} \frac{\partial {v'}}{\partial y} + {w'} \frac{\partial {v'}}{\partial z} \right)\right\rangle }_\text{inertial} +  \underbrace {Ra_F Pr \frac{\partial^2 {\bar \theta}}{\partial z^2}}_{\text{buoyancy}} =  Pr \frac{\partial^4 {\bar w}}{\partial z^4}.  
    \nonumber
\end{eqnarray}
We now omit the inertial and buoyancy terms in equations (\ref{AppB_mom_eq}), (\ref{AppB_zcurl_eq}) and (\ref{AppB_zdoublecurl_eq}) until we estimate them later.
We then let
\begin{equation}
  {\bar w}= {\tilde w(z)} \sin(k_y y), \quad {\bar \omega_z}= {\tilde \omega_z(z)} \sin(k_y y), \quad {\bar u}= {\tilde u(z)} \cos(k_y y),
 \quad {\bar v}= {\tilde v(z)} \cos(k_y y), \quad {\bar \theta}= {\tilde \theta(z)} \sin(k_y y), 
   \label{AppB_tildedef_eq}
\end{equation}
so that (\ref{AppB_zcurl_eq}) gives
\begin{equation}
   -Ta^{1/2} Pr \frac{d {\tilde w}}{d z} = Pr \frac{d^2 {\tilde \omega_z}}{d z^2} + F k_y  \exp \left( \frac{z-1}{H_f} \right) 
   \label{AppB_zcurlb_eq}
\end{equation}
and (\ref{AppB_zdoublecurl_eq}) gives
\begin{equation}
    Ta^{1/2}  \frac{d {\tilde \omega_z}}{d z} =  \frac{d^4 {\tilde w}}{d z^4}. 
    \label{AppB_zdoublecurlb_eq}
\end{equation}
Combining (\ref{AppB_zcurlb_eq}) and (\ref{AppB_zdoublecurlb_eq}) leads to
\begin{equation}
   -Ta Pr \frac{d {\tilde w}}{d z} = Pr \frac{d^5 {\tilde w }}{d z^5} + F Ta^{1/2}k_y  \exp \left( \frac{z-1}{H_f} \right). 
   \label{AppB_w_eq}
\end{equation}
The homogeneous part of the solution for this equation has two solutions which decay into the interior, and two which diverge  there and which must be rejected. 
So
\begin{equation} 
{\tilde w}= a_0 + a_1 \exp \left[ (z-1) Ta^{1/4} \left( \frac{1+i}{\sqrt{2}} \right) \right] +
a_2 \exp \left[ (z-1) Ta^{1/4} \left( \frac{1-i}{\sqrt{2}} \right) \right] + a_3 \exp \left( \frac{z-1}{H_f} \right) 
    \label{AppB_wsol_eq},
\end{equation} 
where
\begin{equation}
    a_3 = \frac{- F k_y H_f}{Ta^{1/2} Pr} \left[ 1 + \frac{1}{H_f^4 Ta} \right]^{-1}.
    \label{AppB_a3sol_eq}
\end{equation}
The boundary conditions at $z=1$ (stress-free) give
\begin{equation}
    {\tilde w} = 0, \quad \frac{d^2 {\tilde w}}{dz^2} =0 , \quad \frac{d^4 {\tilde w}}{dz^4} =0 \ \ {\rm at} \ \ z=1,
    \label{AppB_bc_eq}
\end{equation}
giving
\begin{equation}
    a_0 = - a_3 \left[ 1 + \frac{1}{H_f^4 Ta} \right], \quad a_1 = a_3 \left[ \frac{1}{2 H_f^4 Ta} + \frac{i }{2 H_f^2 Ta ^{1/2}} \right], \quad
    a_2 = a_3 \left[ \frac{1}{2 H_f^4 Ta} - \frac{i }{2 H_f^2 Ta^{1/2}} \right].
    \label{AppB_aisol_eq}
\end{equation}
At the bottom of the top boundary layer, the mean vertical velocity ${\bar w}$ approaches $F k_y H_f \sin (k_y y)/ Pr Ta^{1/2}$. In most runs,
$H_f$ is of the order of the Ekman layer thickness, $Ta^{-1/4}$, for example in run A9, $H_f = (Ta/4)^{-1/4}$. This means that $\bar w \sim O(1)$
when $F \sim O(Ta^{3/4})$. 

Using (\ref{AppB_zdoublecurlb_eq}) we can write the solution for ${\tilde \omega_z}$ as
\begin{eqnarray} 
{\tilde \omega_z} &=& {\hat \omega_z}  - a_1 Ta^{1/4} \frac{1-i}{\sqrt{2}} \exp \left[ (z-1) Ta^{1/4} \left( \frac{1+i}{\sqrt{2}} \right) \right] \nonumber \\
           &-& a_2 Ta^{1/4} \frac{1+i}{\sqrt{2}}  \exp \left[ (z-1) Ta^{1/4} \left( \frac{1-i}{\sqrt{2}} \right) \right] + a_3 
           \frac{1}{H_f^3 Ta^{1/2}} \exp \left( \frac{z-1}{H_f} \right) 
    \label{AppB_omzsol_eq}.
\end{eqnarray} 
where $\hat \omega_z$ is a constant determined by the vorticity in the interior. Using ${\bar \omega_z}=-\partial {\bar u} / \partial y$ and
$\partial {\bar v} / \partial y = - \partial {\bar w} / \partial z$ we can complete the solution for the mean velocity
\begin{eqnarray}
 {\tilde u} &=& \frac{{\hat \omega_z}}{k_y} - \frac{a_1 Ta^{1/4}}{k_y} \frac{1-i}{\sqrt{2}} \exp \left[ (z-1) Ta^{1/4} \left( \frac{1+i}{\sqrt{2}} \right) \right] 
 \nonumber \\
 &-& \frac{a_2 Ta^{1/4}}{k_y} \frac{1+i}{\sqrt{2}}  \exp \left[ (z-1) Ta^{1/4} \left( \frac{1-i}{\sqrt{2}} \right) \right] +  
 \frac{a_3}{H_f^3 k_y Ta^{1/2}} \exp \left( \frac{z-1}{H_f} \right), 
\label{AppB_usol_eq}
\end{eqnarray}
\begin{eqnarray}
 {\tilde v} &=&  \frac{a_1 Ta^{1/4}}{k_y} \frac{1+i}{\sqrt{2}} \exp \left[ (z-1) Ta^{1/4} \left( \frac{1+i}{\sqrt{2}} \right) \right] 
 \nonumber \\
 &+& \frac{a_2 Ta^{1/4}}{k_y} \frac{1-i}{\sqrt{2}}  \exp \left[ (z-1) Ta^{1/4} \left( \frac{1-i}{\sqrt{2}} \right) \right] +  
 \frac{a_3}{H_f k_y } \exp \left( \frac{z-1}{H_f} \right). 
\label{AppB_vsol_eq}
\end{eqnarray}
Using (\ref{AppB_usol_eq}) and defining the jump in ${\bar u}$ across the forced boundary layer as $\Delta {\bar u} = {\bar u}|_{z=1} - {\bar u}|_{int}$,
\begin{equation}
    \Delta {\bar u} = \frac{F  H_f \cos (k_y y)}{\sqrt{2} Pr Ta^{1/4}} \frac{1}{(1+\sqrt{2} H_f Ta^{1/4} + H_f^2 Ta^{1/2}) }.
\label{Deltauestimate_eq}
\end{equation}
Note that with our usual choices $F \sim O(Ta^{3/4})$ and $H_f \sim O(Ta^{-1/4})$, both $\bar u$ and $\bar v$ are $O(Ta^{1/4})$, significantly larger
than the $O(1)$ value of $\bar w$. 

We can now estimate the inertial terms in equation (\ref{AppB_mom_eq}), which were omitted from the boundary layer solution. 
The mean quantities can be estimated from the solution above, 
but the primed quantities depend on the strength of convection i.e. how supercritical the Rayleigh number is and how much the convective turbulence
penetrates into the boundary layer. This is hard to estimate (determining the onset of convection is nontrivial when $F \ne 0$, because the basic state is a function of $y$ as well as $z$). However, from our simulations we find that the fluctuating Reynolds stresses are generally the same order of magnitude as the  Reynolds stresses from the mean quantities, so the same estimate does for both terms for the parameters considered here. To estimate the buoyancy term, we need an estimate for $\bar \theta$ in the boundary layer, which is not simple as nonlinear terms are usually significant in the upper boundary layer. We have to rely on the simulations, which suggest that usually $\bar \theta \sim O(1)$ in the top boundary layer.
 The inertial terms and the buoyancy term can be estimated as terms in the $z$-components of the curl and double curl of the momentum equations (\ref{AppB_zcurl_eq}) and  (\ref{AppB_zdoublecurl_eq}).
 Recalling that when $F \sim O(Ta^{3/4})$ and $H_f \sim O(Ta^{-1/4})$,
 $\tilde w \sim O(1)$, $\omega_z \sim O(Ta^{1/4})$ and $\partial / \partial z \sim O(Ta^{1/4})$, and then ${\bar u},{\bar v},{u'},{v'} \sim O(Ta^{1/4})$
 and ${\bar w}, w' \sim O(1)$. Putting these estimates into equation (\ref{AppB_zcurl_eq}) we see that all the inertial terms are $O(Ta^{1/2})$ but all the other terms are $O(Ta^{3/4})$, so at large $Ta$ the inertial terms are negligible in the boundary layer in this equation. In equation (\ref{AppB_zdoublecurl_eq}) the viscous and Coriolis terms are both $O(Ta)$, but the largest inertial terms are only $O(Ta^{3/4})$
so again the inertial terms are smaller than the viscous and Coriolis terms.
Using the $O(1)$ estimate for $\bar \theta$ we see that the buoyancy term is small compared to the Coriolis and viscous terms in 
(\ref{AppB_zdoublecurl_eq}) provided $Ra_F \ll Ta$, which is the case in most, but not all, of our runs. As expected from this analysis, our estimates of the velocities in the top boundary layer  using this boundary layer theory agree reasonably well with the simulation results.

We can now consider the temperature equation (\ref{mean_temperature_eq}) in the top boundary layer, bearing in mind $\bar\theta \sim O(1)$. Because the boundary condition sets $\partial \theta / \partial z=0 $, $\partial^2 {\bar \theta} / \partial z^2$ is only $O(Ta^{1/4})$ and in fact all the terms in 
equation (\ref{mean_temperature_eq}), including the nonlinear advection terms,
are $O(Ta^{1/4})$. Also, the temperature depends on the matching of the boundary layer to the interior, so it is only possible to solve for the
temperature in the top boundary layer numerically unless $F/Ta^{3/4}$ is small. 

\subsection{Bottom Boundary layer}

The bottom boundary layer near $z=0$  can be treated similarly to the top boundary
layer, but there is no forcing near $z=0$
 so  here, and the boundary conditions are no-slip rather than stress-free. 
 Also, the zonal flow at the bottom of the interior region may not be
 purely sinusoidal, since it may have been affected by the convection in the interior region. The governing equations are (\ref{AppB_zcurl_eq}) and (\ref{AppB_zdoublecurl_eq}) with $F=0$, giving
 \begin{equation}
   -Ta  \frac{\partial {\bar w}}{\partial z} =  \frac{\partial^5 {\bar w }}{\partial z^5}. 
   \label{AppB_w_boteq}
\end{equation}
  Rejecting the two solutions that grow as $z$
increases out of the boundary layer, the solution for $\bar w$ is
\begin{equation} 
{\bar w}= {\hat w}(y) + w_1(y) \exp \left[ -z Ta^{1/4} \left( \frac{1+i}{\sqrt{2}} \right) \right] +
w_2(y) \exp \left[ -z Ta^{1/4} \left( \frac{1-i}{\sqrt{2}} \right) \right] 
    \label{AppB_wsolbot_eq1}.
\end{equation}
Using the boundary conditions
\begin{equation}
 w=0, \qquad \frac{\partial w}{\partial z} =0 \qquad {\rm on} \qquad z=0
 \label{AppB_botbcsw_eq}
\end{equation}
this becomes
\begin{equation} 
{\bar w}= {\hat w}(y) - {\hat w}(y) \frac{1+i}{2} \exp \left[ -z Ta^{1/4} \left( \frac{1+i}{\sqrt{2}} \right) \right] -
{\hat w}(y) \frac{1-i}{2} \exp \left[ -z Ta^{1/4} \left( \frac{1-i}{\sqrt{2}} \right) \right] 
    \label{AppB_wsolbot_eq}.
\end{equation}
The $z$-vorticity is then
\begin{equation} 
{\bar \omega_z}= - \frac{\partial {\hat U}}{\partial y} -
\frac{{\hat w} Ta^{1/4}}{\sqrt{2}} \exp \left[ -z Ta^{1/4} \left( \frac{1+i}{\sqrt{2}} \right) \right] - 
 \frac{{\hat w} Ta^{1/4}}{\sqrt{2}} \exp \left[ -z Ta^{1/4} \left( \frac{1-i}{\sqrt{2}} \right) \right] 
    \label{AppB_omsolbot_eq}
\end{equation} 
using equation (\ref{AppB_zdoublecurlb_eq}) where
$- \partial {\hat U}/{\partial y}$ is the $z$-vorticity at the bottom of the interior region, which matches to the value coming out of the bottom boundary layer. Then the no-slip condition on
$\omega_z$ at $z=0$, $\omega_z=0$ there gives
\begin{equation}
 \frac{\partial {\hat U}}{\partial y} = - {\hat w}(y) \sqrt{2} Ta^{1/4},
 \label{eq_noslip_wu_relation}
\end{equation}
so the vertical velocity (and the other velocity components) can be determined if the zonal flow at the bottom of the interior is known.


\newpage

\begin{table}
\caption{Parameters for the runs performed. The first row of each entry gives the input parameters: Taylor number $Ta$, flux based Rayleigh number $Ra_F$, forcing 
strength in the stable layer $F$, scale depth  of the forcing $H_f$. The parameters controlling the stable layer at the top of the layer are the strength of the stabilising layer $Q$, and the scale depth of the stable layer $H_s$. $z_{neu}$ is the level at which there is neutral stability, the layer being stably stratified for $z>z_{neu}$ and unstably stratified for $z<z_{neu}$. The last two columns of the first row of each entry give the critical Rayleigh number when $F=0$, $Ra_{crit}$, and the corresponding wavenumber, $k_{crit}$. The second row of each entry gives various output quantities. $Ro_{\bar u}$  $Ro_{\bar v \bar w}$ and $Ro_{vw}$ are the Rossby numbers for the mean zonal flow, mean meridional flow and fluctuating meridional flow respectively, see equations (\ref{Ro_ubar_def}) and (\ref{Ro_vw_def}). $\epsilon_{TW}$ measures the departure from thermal wind balance, see (\ref{eps_TW_def}).
$\Delta \bar u_{top}$, $\Delta \bar u_{bot}$, and $\Delta \bar u_{int}$ measure the jumps in mean zonal velocity ${\bar u}$ across the top and bottom boundary layers and the interior respectively (see section IIIA), while the temperature variation ratio ${\bar \theta}_{vr}$ measures the ratio of the sinusoidal parts of $\bar \theta$ at $z=0.2$ to that at $z=0.8$, see 
(\ref{theta_vr_def}). $\tau_2$ is the length of the run in diffusion times, averaging is performed over time $\tau_1=0.5$ to $\tau_2$ except for runs A1 and A14 where $\tau_1=1$ and $3$ respectively.}
\label{table:cases3}
\begin{ruledtabular}
\begin{tabular}{cccccccccc}

 Run  & $Ta$                  & $Ra_F$ & ${F}$ & $H_f$  & $H_s$ &  $Q$ & $z_{neu}$ & $Ra_{crit}$ & $k_{crit}$ \\

  & $Ro_{\bar u}$  & $Ro_{\bar v \bar w}$ & $Ro_{vw}$ & $\epsilon_{TW}$ &  $\Delta \bar u_{top}$ &   $\Delta \bar u_{bot}$ &  $\Delta \bar u_{int}$ & ${\bar \theta}_{vr}$ & $\tau_2$\\
\hline
A1 & $10^{7}$               & $ 0 $                 & 632456                & 0.02515 & 0.03133 & 120    & 0.85  & $3.944\times 10^5$ & 17.81  \\
   & $8.4 \times 10^{-2}$  & $6.8 \times 10^{-3}$ & $6.8 \times 10^{-3}$  &  -      & 45.6    & 369.0  & -1.3  & -0.32             & 1.94   \\
A2 & $10^{7}$               & $ 3 \times 10^{5}$    & 632456                & 0.02515 & 0.03133 & 120    & 0.85  & $3.944\times 10^5$ & 17.81  \\  
   & $6.3 \times 10^{-2}$  & $5.5 \times 10^{-3}$ & $4.9 \times 10^{-2} $ & 0.19   & 44.1   & 247.8  & 58.1  & 0.29              &  1.44  \\
A3 & $10^{7}$               & $ 4 \times 10^{5}$    & 632456                & 0.02515 & 0.03133 & 120    & 0.85  & $3.944\times 10^5$ & 17.81  \\
   & $6.4 \times 10^{-2}$  & $5.1 \times 10^{-3}$ & $6.6 \times 10^{-2} $ & 0.20   & 41.9    & 211.6  & 132.5 & 0.42              & 1.49  \\
A4 & $10^{7}$               & $ 5 \times 10^{5}$    & 632456                & 0.02515 & 0.03133 & 120    & 0.85  & $3.944\times 10^5$ & 17.81  \\
   & $6.8 \times 10^{-2}$  & $4.9 \times 10^{-3}$ & $7.1 \times 10^{-2}$  & 0.21   & 42.6    & 209.6  & 171.7 & 0.45              & 3.89   \\
A5 & $10^{7}$               & $ 6 \times  10^{5}$   & 632456                & 0.02515 & 0.03133 & 120    & 0.85  & $3.944\times 10^5$ & 17.81  \\  
   & $7.0 \times 10^{-2}$  & $4.8 \times 10^{-3}$ & $7.9 \times 10^{-2}$  & 0.22   & 43.1    & 213.1  & 208.6 & 0.46              & 1.20    \\
A6 & $10^{7}$               & $ 10^{7}$             & 632456                & 0.02515 & 0.03133 & 120    & 0.85  & $3.944\times 10^5$ & 17.81  \\ 
   &  $9.8 \times 10^{-2}$ & $4.3 \times 10^{-3}$ & $1.2 \times 10^{-1} $ & 0.41   & 120.4   & 244.5  & 578.9 & 0.34              & 1.27   \\
A7 & $10^{7}$               &  $5 \times 10^5$          & $3 \times 10^5$               & 0.02515 & 0.03133 & 120    & 0.85   & $3.944\times 10^5$ & 17.81  \\
   & $4.0 \times 10^{-2}$  & $2.8 \times 10^{-3}$ & $2.6 \times 10^{-2}$  &  0.063      & 18.6    & 138.6  & 95.4  & 0.25             & 6.25   \\  
A8 & $10^{7}$               & $5 \times 10^{5}$     & 316,228                & 0.02515 &   -     &  0     &  -    & $3.661\times 10^5$ & 17.71  \\ 
   & $3.4 \times 10^{-2}$  & $3.5 \times 10^{-3}$ & $4.0 \times 10^{-3}$  & 0.0053  & 22.7    & 187.3  & -47.5 & 1.02               & 2.29   \\
A9 & $10^{7}$               & $ 5 \times 10^{5}$    & 316,228                & 0.02515 & 0.02831 & 200    & 0.85  & $3.963\times 10^5$ & 17.84  \\ 
   & $4.1 \times 10^{-2}$  & $2.7 \times 10^{-3}$ & $3.4 \times 10^{-2}$  & 0.098  & 20.8    & 131.0  & 123.1 & 0.27             & 4.01   \\
A10 & $10^{7}$               & $ 5 \times 10^{5}$    & 632456                & 0.02515 & 0.02831 & 200    & 0.85  & $3.963\times 10^5$ & 17.84  \\  
   & $7.4 \times 10^{-2}$  & $4.8 \times 10^{-3}$ & $8.9 \times 10^{-2}$  & 0.25  & 48.3    & 216.3  & 231.9 & 0.43              & 4.17    \\
A11 & $10^{7}$               & $ 5 \times  10^{5}$   & 632456                & 0.02515 & 0.01887 & 200    & 0.90  & $3.808\times 10^5$ & 17.78  \\
   & $6.3 \times 10^{-2}$  & $5.2 \times 10^{-3}$ & $6.0 \times 10^{-2}$  & 0.20  & 43.1    & 221.9  & 109.0 & 0.39              & 2.53   \\
A12 & $10^{7}$               & $ 5 \times 10^{5}$    & 632456                & 0.02515 & 0.01654 & 423    & 0.90  & $3.832\times 10^5$ & 17.81  \\
   & $7.1 \times 10^{-2}$  & $4.8 \times 10^{-3}$ & $7.6 \times 10^{-2}$  & 0.25  & 48.4    & 212.2  & 208.4 & 0.43              & 1.80    \\
A13 & $10^{7}$               & $ 5 \times  10^{5}$   & 632456                & 0.02515 & 0.00625 & 600    & 0.96  & $3.716\times 10^5$ & 17.74  \\
   & $7.1 \times 10^{-2}$  & $6.3 \times 10^{-3}$ & $2.9 \times 10^{-2}$  & 0.076  & 43.7    & 339.3  & -47.0 & 2.89               & 4.99   \\
A14 & $10^{7}$               & $ 0 $                 & 10,000                 & 0.02515 & 0.03133 & 120    &  0.85 & $3.944\times 10^5$ & 17.81  \\
   & $1.3 \times 10^{-3}$  & $1.1 \times 10^{-4}$ & $1.1 \times 10^{-4}$  & -       & 0.71    & 5.9    &  -0.04 & 0.28              & 3.99    \\
B1 & $10^{8}$ & $ 10^{6}$          & 632456 & 0.02515 & 0.03133 & 120 & 0.85 & $1.827\times 10^6$ & 26.53   \\
 & $1.6 \times 10^{-2}$ & $8.0 \times 10^{-4}$ & $3.5 \times 10^{-3}$ & 0.0073 & 10.7 & 214.0 & 41.8 & 0.070 & 4.36 \\ 
B2 & $10^{8}$ & $ 10^{6}$          & $2 \times 10^{6}$ & 0.01414 & 0.03133 & 120 & 0.85 & $1.827\times 10^6$ & 26.53  \\  
 & $2.7 \times 10^{-2}$ & $1.5 \times 10^{-3}$ & $9.3 \times 10^{-3}$ & 0.017 & 37.2 & 355.0 & 73.9 & 0.088 & 6.20 \\  
B3 & $10^{8}$ &  $ 5 \times 10^{7}$ &  $4 \times 10^{4}$ & 0.02 & 0.02 & 300 & 0.886  & $1.784\times 10^6$ & 26.51 \\   
  & $3.0 \times 10^{-3}$ & $1.3 \times 10^{-4}$ & $1.8 \times 10^{-2}$  & 0.082 &  3.9 & 12.5 & 65.9 & 0.73 & 11.3 \\
B4 & $10^{8}$ &  $ 5 \times 10^{7}$ &  $5 \times 10^{5}$ & 0.02 & 0.02 & 300 & 0.886  & $1.784\times 10^6$ & 26.51 \\   
  & $1.0 \times 10^{-2}$ & $4.2 \times 10^{-4}$ & $2.3 \times 10^{-2}$  & 0.070 &  31.3 & 101.9 & 193.1 & 0.18 & 14.75 \\
B5 & $10^{8}$ &  $ 5 \times 10^{7}$ &  $ 10^{6}$ & 0.02 & 0.02 & 300 & 0.886  & $1.784\times 10^6$ & 26.51 \\   
  & $1.7 \times 10^{-2}$ & $7.9 \times 10^{-4}$ & $2.9 \times 10^{-2}$  & 0.062 &  45.4 & 171.8 & 291.1 & 0.11 & 11.95 \\
C1 & $10^{9}$ & $5 \times 10^{8}$   & $2 \times 10^{5}$ & 0.01 & 0.00815 & 460 &  0.95 & $8.054\times 10^6$ & 39.24  \\
  & $9.8 \times 10^{-4}$ & $2.6 \times 10^{-5}$ & $1.3 \times 10^{-2}$ & 0.072 & 1.7 & 6.0 & 68.3 & 0.70 &  2.78  \\      
 \end{tabular}
 \end{ruledtabular}
\end{table}

\clearpage

%

\end{document}